\documentclass[preprint,12pt]{elsarticle}
\usepackage[a4paper,margin=2.5cm]{geometry} 
\usepackage{xcolor,array,enumitem,makecell}
\usepackage{newtxtext,newtxmath}
\usepackage{pgfplotstable,siunitx,booktabs,colortbl}
\pgfplotsset{compat=1.8}

\usepackage{hyperref}

\journal{Advanced Materials}

\begin{document}

\begin{frontmatter}



\title{AI-Driven Defect Engineering for Advanced Thermoelectric Materials}

\cortext[cor1]{Equal contribution}
\affiliation[1]{organization={Quantum Measurement Group, MIT},
            city={Cambridge},
            state={MA 02139},
            country={USA}}
\affiliation[2]{organization={Department of Nuclear Science $\&$ Engineering, MIT},
           city={Cambridge},
            state={MA 02139},
            country={USA}}
\affiliation[3]{organization={Center for Computational Science $\&$ Engineering, MIT},
           city={Cambridge},
            state={MA 02139},
            country={USA}}
\affiliation[4]{organization={Department of Materials Science $\&$ Engineering, MIT},
           city={Cambridge},
            state={MA 02139},
            country={USA}}
\affiliation[5]{organization={Frontier Research Institute for Interdisciplinary Sciences, Tohoku University},
           city={Sendai, 980-8578},
            country={Japan}}
\affiliation[6]{organization={Linac Coherent Light Source, SLAC National Accelerator Laboratory}, 
        city={Menlo Park}, 
        state={CA 94025}, 
        country={USA}}
\affiliation[7]{organization={Stanford Institute for Materials and Energy Sciences, Stanford University}, 
        city={Stanford}, 
        state={CA 94025}, 
        country={USA}}
\affiliation[8]{organization={Department of Chemistry, MIT},
           city={Cambridge},
            state={MA 02139},
            country={USA}}
\affiliation[9]{organization={Neutron Scattering Division, Oak Ridge National Laboratory}, 
        city={Oak Ridge}, 
        state={TN 37831}, 
        country={USA}}
\author[1,2]{Chu-Liang Fu\corref{cor1}}
\ead{clfu2357@mit.edu}
\author[1,3,4]{Mouyang Cheng\corref{cor1}}
\author[5]{Nguyen Tuan Hung\corref{cor1}}
\author[1,2]{Eunbi Rha}
\author[6,7]{Zhantao Chen}
\author[1,8]{Ryotaro Okabe}
\author[1,2]{Denisse Córdova Carrizales}
\author[1,2]{Manasi Mandal}
\author[9]{Yongqiang Cheng}
\author[1,2]{Mingda Li}
\ead{mingda@mit.edu}

\begin{abstract}
Thermoelectric materials offer a promising pathway to directly convert waste heat to electricity. However, achieving high performance remains challenging due to intrinsic trade-offs between electrical conductivity, the Seebeck coefficient, and thermal conductivity, which are further complicated by the presence of defects. This review explores how artificial intelligence (AI) and machine learning (ML) are transforming thermoelectric materials design. Advanced ML approaches including deep neural networks, graph-based models, and transformer architectures, integrated with high-throughput simulations and growing databases, effectively capture structure-property relationships in a complex multiscale defect space and overcome the curse of dimensionality. This review discusses AI-enhanced defect engineering strategies such as composition optimization, entropy and dislocation engineering, and grain boundary design, along with emerging inverse design techniques for generating materials with targeted properties. Finally, it outlines future opportunities in novel physics mechanisms and sustainability, highlighting the critical role of AI in accelerating the discovery of thermoelectric materials.
\end{abstract}

\end{frontmatter}

\section{Introduction}
\label{sec:intro}
Energy materials play an indispensable role in addressing global challenges and also foster economic growth and technological innovation. For instance, perovskites solar cells have achieved high photovoltaic efficiency \cite{hodes2013perovskite,tonui2018perovskites,shi2018perovskites}, solid-state electrolytes have enhanced battery performance and safety \cite{fan2018recent,zhao2020designing}, and hydrogen storage alloys \cite{sandrock1999panoramic,marques2021review} have improved the feasibility of hydrogen-based energy systems. The development of novel advanced materials has opened new pathways for the future of sustainable energy.

However, the design of new energy materials faces tremendous challenges. The first difficulty lies in the structural complexity \cite{snyder2008complex,guan2017complex}, which is aggregated when considering the effects of defects \cite{bai2018defect,zhang2020defect,zheng2021defect,wu2024defect}. While defects provide materials scientists a tuning degree of freedom to potentially enhance performance, the inclusion of defects significantly increases the dimensionality of the materials' design space. This complexity is further amplified by the multiscale nature of defects, posing a second challenge for comprehensive defect modeling  \cite{wu2024defect,zhang2021multiscale,li2023multi}. Third, optimizing energy materials often requires a simultaneous balance of multiple material properties, adding yet another layer of complexity \cite{snyder2008complex}. Fourth, sustainability has emerged as a pressing design factor, prompting researchers to integrate eco-friendly practices, life-cycle assessments, and environmental impact evaluations into materials design \cite{zhou2009multi,yao2023machine,boonkird2023sustainability}. All factors collectively create an extremely high-dimensional design space, making traditional trial-and-error methods -- whether experimental or computational -- increasingly impractical to search for even a small region in the design space. 

Thermoelectric (TE) materials are an important class of energy materials to address future energy challenges \cite{snyder2008complex,sootsman2009new,rowe2018thermoelectrics}. TE materials directly convert heat into electricity or vice versa, removing heat via electrical current, without moving parts, making them highly attractive for applications in waste heat recovery \cite{orr2016review,burnete2022review,ovik2016review}, solid-state cooling \cite{mao2021thermoelectric,qin2022solid}, renewable energy systems \cite{elsheikh2014review,zheng2014review}, and IoT sensors~\cite{haras2018thermoelectricity}. However, achieving a high TE figure-of-merit ($zT$) requires simultaneous optimization of electrical conductivity, Seebeck coefficient, and thermal conductivity -- properties that are interdependent and often constrained by conflicting trade-offs. Moreover, these properties are sensitive to the defect structure of materials that span multiple scales \cite{snyder2008complex,zhu2017compromise,wu2024defect}. 

In the past few decades, many efforts have been dedicated to addressing this challenge, such as using low-dimensional materials~\cite{hicks1993effect,hicks1993thermoelectric,hung2016quantum}, the convergence of electronic bands~\cite{t2019thermoelectric}, and resonant-state doping~\cite{heremans2008enhancement}. Combined with modern computational methods such as first-principle calculations and high-throughput screening ~\cite{hung2019designing,deng2024high,zhou2016first}, high values of maximum $zT$ of 2 (i.e. about 15\% efficiency of energy conversion at cold-side temperature $T_{c}=300$K and hot-side temperature $T_{h}=560$K) were reported in some of the TE materials, such as SnSe~\cite{chang20183d,zhao2014ultralow}, Cu$_2$Se~\cite{olvera2017partial,zhong2014high}, PbTe~\cite{tan2016non}, GeTe~\cite{li2018low}, or Bi$_2$Te$_3$/Sb$_2$Te$_3$ superlattice~\cite{venkatasubramanian2001thin}. However, the complex defect configuration may have hindered the study of the underlying mechanism and complicates the scalability. Although thermoelectric materials with $zT = 2$ approach the state-of-the-art efficiency (10\% $\sim$15\%), they still lag behind other technologies, such as single-crystal silicon solar cells ($\sim$ 26.1\%) and III-V multi-junction solar cells with concentrators ($\sim$47.6\%). To make TE more competitive compared to other technologies, such as photovoltaic, the TE device should operate with 30\% efficiency, which corresponds to $zT \sim 10$~\cite{tang2024high}. 


In recent years, machine learning (ML) has led to a paradigm shift in accelerating the discovery and design of materials \cite{gubernatis2018machine,moosavi2020role,pollice2021data,choudhary2022recent}. By leveraging extensive datasets from experiments and simulations, ML algorithms enable the prediction of critical structure-property relationships in materials, including thermal transport \cite{wei2022perspective,ju2020designing}, electrical transport \cite {zhuo2018predicting,li2019electronic,chandrasekaran2019solving}, magnetism prediction \cite{merker2022machine}, vibrational properties \cite{hanYQ2024ai} and optical properties \cite{hung2024universal}, thereby accelerating the design and development of energy materials. Furthermore, ML facilitates the generation of candidate materials under specific conditions \cite{sanchez2018inverse,zeni2025mattergen}, and supports autonomous materials discovery for synthesis \cite{szymanski2023autonomous}. Moreover, it enhances characterization techniques by improving the analysis of spectroscopy methods such as neutron and X-ray scattering \cite{chen2021machine}, Raman spectroscopy \cite{qi2023recent}, electron microscopy \cite{kalinin2022machine}, and electrochemical impedance spectroscopy \cite{zhang2020identifying}, etc. One of ML’s key strengths lies in its ability to handle multiscale simulations of energy materials \cite{mortazavi2024recent}. For instance, ML-driven surrogate models can bridge the gap between first-principles electronic structure calculations and mesoscale transport simulations, providing a holistic understanding of material behavior across scales \cite{zhang2018deep,hu2021perspective,chen2021applying}. Utilizing such data-driven methods helps mitigate the long-standing ``curse of dimensionality" in energy material design. Moreover, with sustainability becoming more important in materials science, ML offers powerful tools to promote environmentally conscious innovation \cite{yao2023machine}, such as optimizing battery charging and discharging cycles \cite{wei2014novel} and predicting materials degradation \cite{hsu2020using}. The capability of ML for efficient optimization enables researchers to design high-performance materials while minimizing environmental impact. 

This review mainly focuses on how ML transforms TE materials design by addressing structural complexity. Firstly, it discusses the background on the design of TE materials and current bottlenecks. Then, potential solutions will be presented to address the bottleneck through the recent progress of advanced ML paradigm, including the architectures and algorithms, hardware, dataset, and descriptors. The next topic will be how to utilize these ML approaches along with various defect engineering techniques in the design of TE materials, such as doping strategy, entropy engineering, dislocation and interface engineering. Finally, several future directions for optimizing TE materials are proposed, including additional opportunities to apply ML techniques such as inverse design, topological TE, and sustainability.

\section{Background and Bottleneck in Thermoelectric Materials Design}
\begin{figure}[t]
    \centering
    \includegraphics[width=0.8\linewidth]{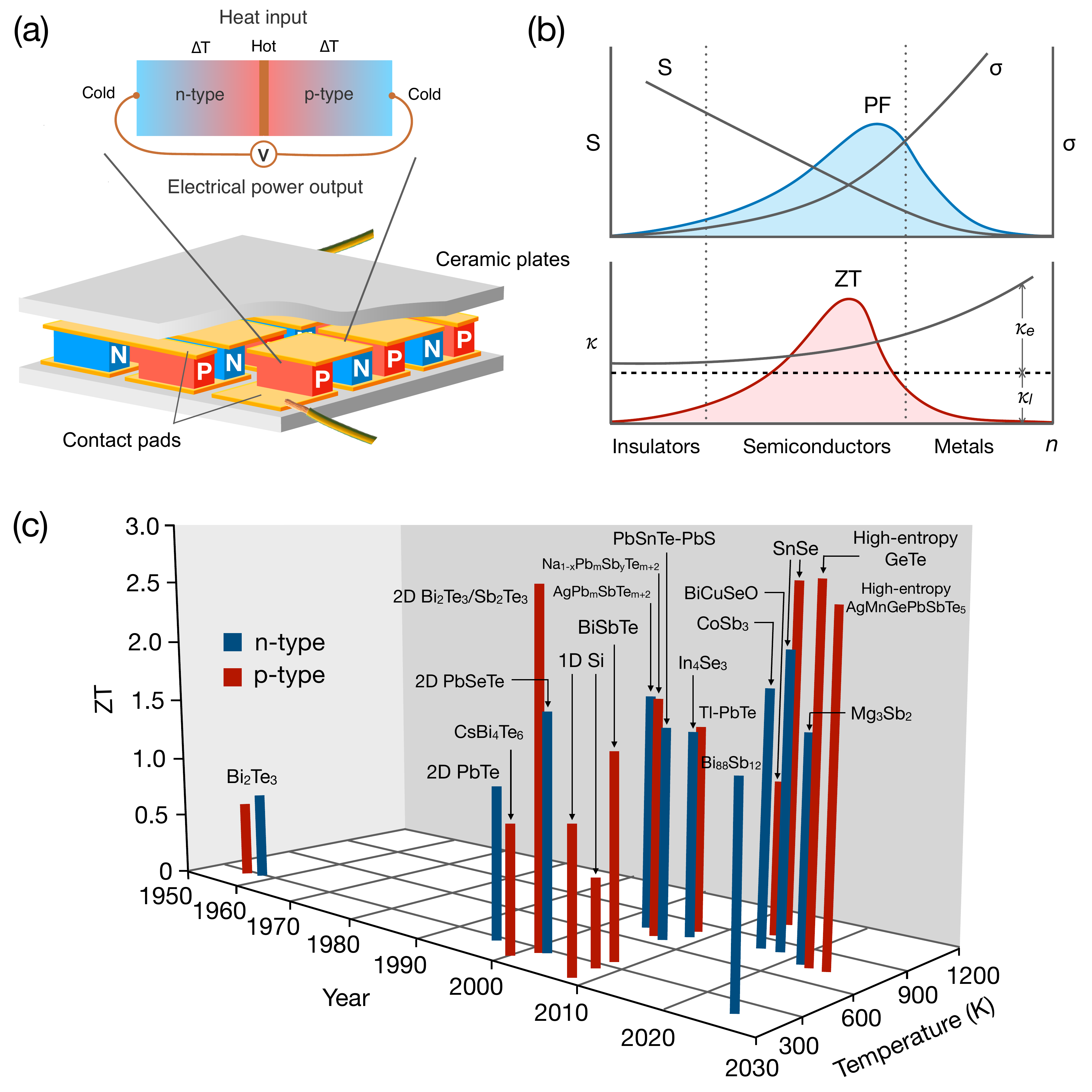}
    \caption{\textbf{TE device and performance. (a)} A schematic of a TE device showing many individual thermocouples that are connected in electrical series. \textbf{(b)} Illustration of the variation of Seebeck coefficient $S$, electrical conductivity $\sigma$, total thermal conductivity $\kappa=\kappa_e+\kappa_l$, where $\kappa_e$ and $\kappa_l$ are the electron and lattice thermal conductivities, power factor $\text{PF}=S^2\sigma$, and figure of merit $zT=\text{PF}\times T/\kappa$ as a function of the carrier concentration. \textbf{(c)} $zT$ as a function of temperature and year. Figure adapted and updated from Ref ~\cite{hung2021origin} with new updates until 2025.}
    \label{fig:zT}
\end{figure}

 TE energy conversion is based on the Seebeck effect in which a temperature gradient induces a voltage across a material. Discovered in 1821 by Thomas Johann Seebeck~\cite{macdonald2006thermoelectricity}, this effect enables direct heat-to-electricity conversion without moving parts. As shown in Fig.~\ref{fig:zT}(a), 
heating the junction of the $n$-type and $p$-type semiconductors (or a thermocouple) generates a voltage between their free ends. The Seebeck coefficient, $S$, is defined by how much voltage, $\nabla V$, is generated by the temperature gradient, $\nabla T$, as~\cite{macdonald2006thermoelectricity} 
\begin{equation}
    S = -\frac{\nabla V}{\nabla T}.
\end{equation}

Based on the Seebeck effect, a TE device consists of multiple thermocouples connected in electrical series and thermal parallel, as shown in Fig.~\ref{fig:zT}(a). This configuration enhances voltage output while maintaining efficient heat flow. There are two parameters to evaluate the performance of a TE device: the maximum output power density, $Q$, and the maximum TE efficiency, $\eta$. When the heat source is relatively free, such as solar heat or waste heat, the minimum cost of generating TE power is achieved by operating at a maximum of $Q$. On the other hand, when the heat source is expensive, such as fossil fuel combustion, maximum $\eta$ is desired to reduce the cost to generate power~\cite{hung2016quantum,liu2015n,liu2016importance}. The maximum of $Q$ and $\eta$ are defined by~\cite{hung2021origin}
\begin{equation}\label{eq:PF}
    Q=\frac{1}{4L}\text{PF}(\Delta T)^2,
\end{equation}
and
\begin{equation}\label{eq:zT}
    \eta=\frac{T_h-T_c}{T_h}\times \frac{\sqrt{1+zT}-1}{\sqrt{1+zT}+T_c/T_h},
\end{equation}
where $L$ is the leg length of the thermocouple, $\Delta T = T_h-T_c$ is the temperature difference between the hot side $T_h$ and the cool side $T_c$, and $T=(T_c+T_h)/2$ is the average temperature. The PF and $zT$ are the TE power factor and dimensionless figure of merit, which are defined by
\begin{equation}
    \text{PF}=S^2\sigma,
\end{equation}
and
\begin{equation}\label{zT}
    zT = \frac{\sigma S^2 T}{\kappa},
\end{equation}
where $\sigma$ is the electrical conductivity,  $\kappa=\kappa_e+\kappa_l$ the total thermal conductivity composed of the electronic ($\kappa_e$) and lattice phonon ($\kappa_l$) contributions. 

Equations~\eqref{eq:PF} and~\eqref{eq:zT} show that achieving a high TE performance (i.e. $Q$ or $\eta$) requires maximizing PF, $zT$ or both. 
However, such an optimization is fundamentally challenging since $S$, $\sigma$, and $\kappa$ are mutually dependent. As shown in Fig.~\ref{fig:zT}(b), for metals, only electrons near the Fermi energy $E_F$ contribute to $S$. Thus, according to the Mott formula, $S$ should be proportional to $k_BT/E_F$~\cite{ioffe1957semiconductor}, where $k_B$ is the Boltzmann constant. For a monovalent metal with $k_BT < E_F$, $S$ often has a small value around a few \SI{}{\micro\volt\per\kelvin}. Moreover, metals often have a high $\kappa$, which contains the contribution of $\kappa_e$, known as the Wiedemann–Franz law, in which $\kappa_e=LT\sigma$, where $L$ is the Lorenz number. In contrast, an insulator has larger $S$ since the maximum $S$ is proportional to the energy band gap~\cite{hung2015diameter}, but it has a small $\sigma$. This trade-off problem can be considered the first bottleneck that limits TE performance. 

The breakthrough in overcoming the bottleneck came in 1949 when Ioffe discovered that heavily doped semiconductors~\cite{ioffe1957semiconductor}, such as bismuth telluride, lead telluride, and silicon-germanium alloys, could achieve a high $zT$ value. Unlike insulators and metals, semiconductors can have optimized values of the PF and $zT$, as shown in Fig.~\ref{fig:zT} (b). This advancement in doped semiconductors partially solves the trade-off problem, enabling the practical applications of TE materials. Soon after, the first TE generator was launched into Earth's orbit on a U.S. space mission in 1961. 

\begin{figure}
    \centering
    \includegraphics[width=0.85\linewidth]{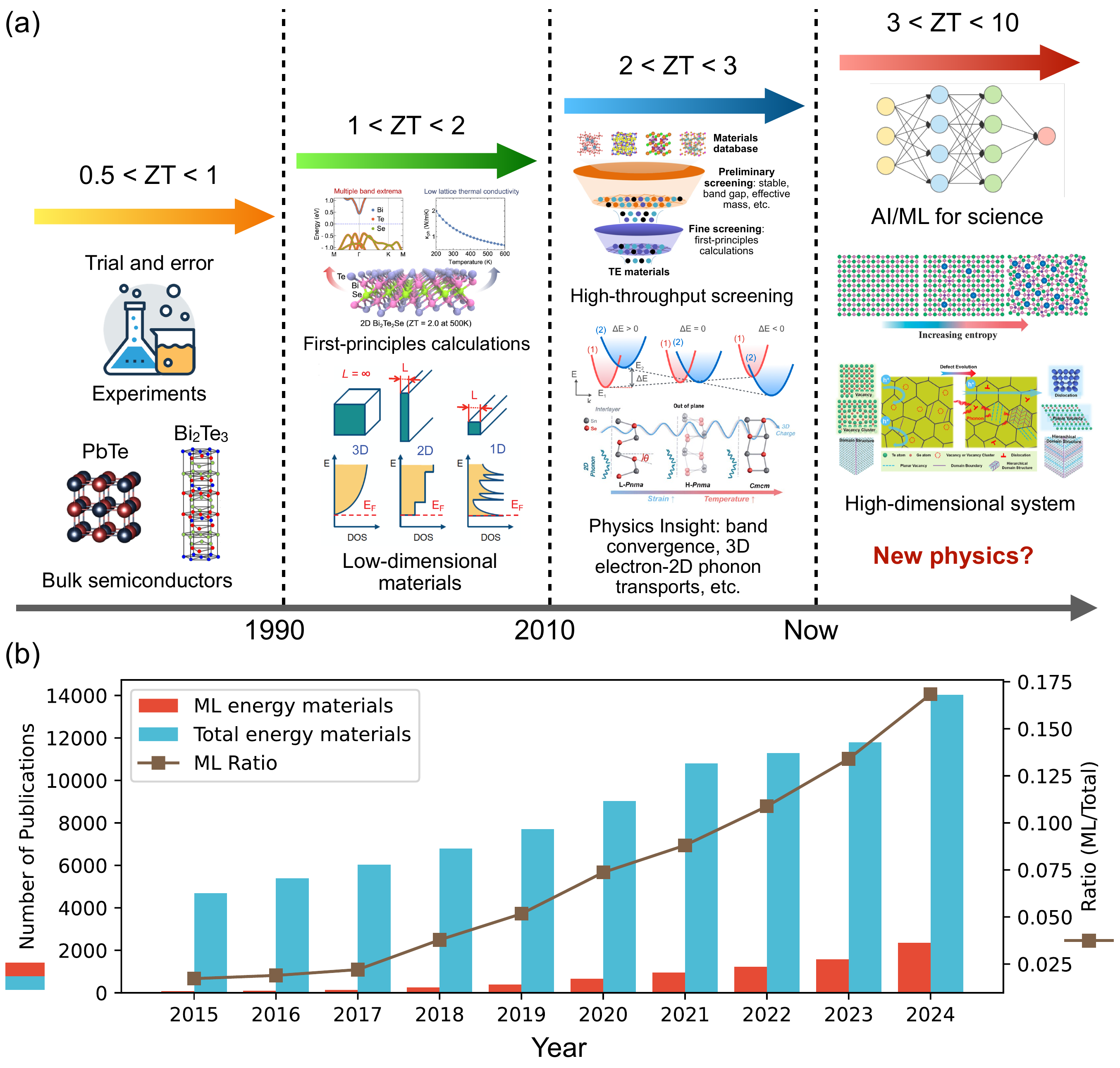}
    \caption{\textbf{Development stages of TE materials and the broader energy materials research.} \textbf{(a)} Evolution of $zT$ value over time in the search for TE materials.
    Powered by computational methods and advanced AI techniques, the paradigm has shifted from experimental trial and error to high-throughput calculations and AI-driven discovery, leading to new TE materials with higher and higher figures of merit $zT$. Reproduced with permission of Refs.~\cite{t2019thermoelectric,hung2019designing,su2022high,jiang2022evolution,jiang2021high}. \textbf{(b)} Evolutionary trend of energy materials research since 2015. The publications presented in this figure are retrieved from the Web of Science database with keywords related to automatic energy materials design and discovery. 
    Publications with ML energy materials are filtered based on additional keywords with various AI-based techniques.}
    \label{fig:TE}
\end{figure}

After 1960, despite extensive research efforts in TE semiconductors, improvements in the TE efficiency remained limited\textemdash not exceeding 10\% efficiency (or $zT < 1$) for decades, as shown in Fig.~\ref{fig:zT}(c). Thus, TE materials were mostly only used in space exploration before the 1990s. To utilize TE materials for other applications in energy harvesting and waste heat recovery, a new benchmark for $zT$ was established. Therefore, a long-standing goal is to discover materials with $zT = 2$~\cite{majumdar2004thermoelectricity}. This goal can be considered the second bottleneck in the TE field, as shown in Fig.~\ref{fig:TE}(a).

One of the solutions for the second bottleneck was proposed in 1993 by Hicks and Dresselhaus, who theoretically showed that reducing the dimensionality of TE materials could dramatically improve their performance~\cite{hicks1993effect,hicks1993thermoelectric}. Their studies suggested that in low-dimensional (1D and 2D) materials the confinement effect can lead to van Hove singularities in the electronic density of states, which can enhance carrier concentration near the bottom (or top) of the conduction (or valence) band and thereby enhance the PF. A more precise condition to enhance the PF of the 1D and 2D semiconductors was elucidated by Hung \textit{et al.}~\cite{hung2016quantum,hung2021origin}. Low-dimensional materials may also have suppressed lattice thermal conductivity $\kappa_l$ due to enhanced phonon-boundary scattering when the diameter or thickness is smaller than the mean free path of phonons (in the range of 100 nm to 1 $\mu$m). Thus, low-dimensional materials are expected to have a high $zT$ value, as shown in Fig.~\ref{fig:zT}(c). In 2001, Venkatasubramanian \textit{et al.}~\cite{venkatasubramanian2001thin} reported a room-temperature $zT$ of about 2.4 for 2D superlattices of Bi$_2$Te$_3$/Sb$_2$Te$_3$.

In 2008, Chen and Ren's teams introduced nanocomposite TE materials as an alternative to low-dimensional materials. They created the nanocomposite powder by ball milling and hot-pressing TE materials into dense 3D pellets. These bulk nanostructured materials offer better scalability than the 1D and 2D structures for TE applications. The nanostructured bismuth-antimony-telluride (BiSbTe) has reported about $zT=1.4$ at room temperature, a 40\% improvement over its bulk counterparts~\cite{poudel2008high}. This breakthrough has revived the research interest in 3D TE materials until today, as shown in Fig.~\ref{fig:zT}(c). Combined with first-principles calculations and high-throughput screening~\cite{hung2019designing,deng2024high,zhou2016first}, several mechanisms to improve the $zT$ values have been exposed. For example, band convergence refers to the alignment or convergence of multiple energy sub-bands of an electronic structure, resulting in enhanced PF and $zT$~\cite{pei2011convergence,t2019thermoelectric} or simultaneously 3D charge and 2D phonon confirmed transport, leading to high out-of-plane $zT$ in SnSe crystals~\cite{chang20183d}. These advances contribute to the enhancement of $zT$, in which some TE materials reported with $2 < zT < 3$, as shown in Fig.~\ref{fig:TE}(a). 

Recently, high-entropy materials (i.e., materials formed by mixing equal or relatively large proportions of multiple elements) have attracted attention in TE research due to their ultralow lattice thermal conductivity $\kappa_l$. The distorted lattices in the high-entropy material lead to a large intrinsic local strain, providing strong phonon scattering to reduce $\kappa_l$ and thereby reaching higher $zT$. Jiang \textit{et al.}~\cite{jiang2022high} reported that high-entropy GeTe-based material has a maximum $zT$ of 2.7 at 750K. Besides high-entropy materials, $\kappa_l$ can also be suppressed through point defects, dislocations, and grain boundary engineering ~\cite{novitskii2024defect,kuo2018grain}. However, the increased degrees of freedom simultaneously enlarge the high-dimensional design space and complicate the design process of TE materials. To push TE performance further, the next goal could be to set a target $zT$ up to 10, as shown in Fig.~\ref{fig:TE}(a). 

\section{Machine Learning for Materials Design}
ML has emerged as a powerful tool in materials science, offering a data-driven paradigm to accelerate materials discovery and design. The core workflow of ML begins with collecting data from experiments or simulations, followed by designing a tailored learning model to capture the relationships between materials' structures, properties, processing, and performance. These relationships are often too complex to be seen in a straightforward manner. Once the learning model is designed, it is trained using the collected data by minimizing a specific loss function. This process effectively constrains the model to a particular variable space. Subsequently, the model is evaluated to ensure it produces reliable predictions. Several reviews have already outlined the foundations of ML in energy materials design \cite{gu2019machine,chen2020critical,liu2021machine,wang2020simulation,lu2021computational,kim2024review} with some focused specifically on TE materials \cite{wang2020machine,antunes2022machine,wang2023critical}. Our survey indicates a rapid growth in ML-related publications on energy materials since 2015 (Fig.\,\ref{fig:TE}(b)). However, beyond the growing interest, it is essential to explore why ML can be transformative in this field. This section highlights key advancements that have enabled ML to support energy materials studies: advancements in ML algorithms and architectures (Sec. 3.1), the rise of powerful computational hardware (Sec. 3.2), the recent accumulation of extensive datasets from high-throughput simulations and experiments (Sec. 3.3), and the development of better representation and descriptors for materials structures (Sec. 3.4).

\subsection{ML architectures and algorithms for complex energy materials}
ML algorithms, particularly deep learning architectures based on neural networks, have advanced rapidly over the past few decades \cite{jordan2015machine,lecun2015deep}. Before the advent of neural networks, a fundamental challenge in high-dimensional data modeling was the \textit{curse of dimensionality}, where computational requirements grow exponentially with the number of input variables. Traditional approaches, such as polynomial regression or kernel methods, often fail to resolve this issue, making it challenging to construct accurate mappings for high-dimensional structure-property relationships. In TE materials design, where the dimensionality of the design space continues to expand, this problem becomes particularly pronounced. 
Deep neural networks (DNNs) address this issue by leveraging hierarchical feature extraction to approximate highly complex functions, effectively mitigating the curse of dimensionality \cite{pinkus1999approximation,yarotsky2017error,bach2017breaking,suzuki2018adaptivity,weinan2022mathematical}. 
For this reason, DNNs have gained popularity in a wide range of scientific and engineering domains, including energy materials design. In particular, DNNs enable efficient exploration of vast compositional, structural, and defect spaces and facilitate predictive modeling of TE performance. The following discussion provides a brief introduction to some deep learning techniques relevant to energy materials design. A more general perspective of AI methods for materials design can be found in Ref.\,\cite{cheng2025ai}.

The foundation of deep learning in high-dimensional modeling was established by early architectures like multi-layer perceptrons (MLPs), which can approximate arbitrary continuous functions according to the universal function approximation theorems.
Based on MLPs, convolutional neural networks (CNNs) introduced a pivotal breakthrough by incorporating spatially localized receptive fields and weight-sharing mechanisms, allowing them to learn hierarchical feature representations with significantly fewer parameters\cite{lecun1998gradient}. This innovation proved particularly valuable in domains such as image processing and structured data analysis, where spatial coherence is a critical factor.
However, both MLPs and early CNNs suffer from the vanishing gradient problem, where the gradients of learnable weights shrink exponentially as the depth of the neural network increases. This could degrade the performance of DNNs when more layers are added \cite{bengio1994learning,glorot2010understanding,he2015convolutional}. 

A major breakthrough in deep learning architecture was the introduction of the residual network (ResNet)\cite{he2016deep}, which overcame the vanishing gradient issues in deep models. 
Specifically, ResNet decomposes a non-linear function $H(x)$ of each layer into $H(x)=f(x)+x$ where $x$ is the identity mapping from the previous layer, and $f(x)=H(x)-x$ is the \textit{residual} component of the nonlinear function parametrized by learnable weights.
ResNet is typically realized through the skip-connection mechanism, where the input to a given layer is directly added to the output of a deeper layer. Such implementation allows models to scale to hundreds of layers without degradation in performance. Due to this advantage, ResNet blocks with residual connections have become a fundamental component in more recent modern DNN architectures, including Stable Diffusion and ChatGPT.

The architecture of ResNet not only empowers modern advanced neural networks but also offers valuable insights into input representation.
The essence of a residual connection is to decompose the input information into two components: a baseline signal that is straightforward to learn (e.g., the identity mapping $x$), and a more intricate, nonlinear residual component that captures nuanced variations (e.g., the residual function $f(x)$). This concept is particularly relevant in designing energy materials like TE materials, where the chemical design space becomes more complicated due to the presence of defects (Sec. 4).

Beyond MLP and CNN-based architectures, transformer-based networks have emerged as another dominant paradigm, particularly in sequence modeling and high-dimensional relational learning.
One key innovation in transformer architectures is the self-attention mechanism \cite{vaswani2017attention}, which dynamically weights interactions between all input elements and efficiently captures long-range dependencies.
By encoding material representations as sequential inputs and capturing their intricate relations, the transformer architecture with self-attention blocks could provide insight for optimizing energy materials like TE materials (Sec. 4).

Apart from the deep learning architectures mentioned above, which primarily focus on supervised learning, deep generative models trained through unsupervised learning and self-supervised learning offer promising opportunities for energy materials design. The generative models, including variational autoencoders (VAEs)\cite{kingma2013auto}, generative adversarial networks (GANs)\cite{goodfellow2014generative}, diffusion models\cite{ho2020denoising} and large language models (LLMs) \cite{radford2019language} hold immense potential for inverse materials design, where new materials are discovered with tailored properties. 
By leveraging probabilistic modeling, these models can generate realistic material candidates while incorporating stability constraints and enabling \textit{conditional} generation of materials with targeted properties.

\subsection{AI hardware for materials science}
Another critical factor facilitating the data-driven revolution in materials science is the rapid advancement of hardware acceleration for high-performance computing (HPC) during recent years \cite{dally2021evolution,khan2021advancements}. Hardware acceleration is the use of computer hardware, such as graphics processing units (GPUs), to conduct specific functions more efficiently compared to running the same jobs on general-purpose central processing units (CPUs). The advancement in hardware acceleration has made deep learning practical for AI, transforming it from a theoretical concept into a scalable and efficient tool \cite{chellapilla2006high,krizhevsky2012imagenet}. Specialized hardware like GPU, tensor processing units (TPUs) \cite{cass2019taking}, and AI accelerators enable the rapid training and deployment of deep learning models by optimizing massive parallel computations, tensor operations, and backpropagation, making AI breakthroughs feasible across various domains.

Many computational tasks in materials discovery are inherently parallel and well-suited for GPU-based architectures, making hardware acceleration a crucial enabler of modern materials research \cite{andreoni2000new,vondrous2014parallel,zhang2024gpu}. Modern materials research increasingly relies on computational approaches, including high-throughput materials screening for targeted properties such as TE-related properties \cite{chen2016understanding}, atomic simulations \cite{zhang2018deep}, and ML-driven discovery \cite{sanchez2018inverse}. These tasks require massive computational power, while the computational workflows often involve repetitive and independent calculations performed across vast datasets, making them particularly well-suited for parallel architectures. The highly parallel nature of these problems naturally fits GPU-based hardware acceleration, as GPUs are optimized for executing thousands of threads simultaneously. Unlike traditional CPUs, which are designed for sequential processing and general-purpose computing, GPU leverages a Single-Instruction, Multiple-Threads (SIMT) execution model, allowing multiple threads to process data concurrently with minimal overhead \cite{weinzierl2022principles}. This makes GPUs particularly advantageous for scientific computing tasks involving linear algebra operations, tensor computations, and large-scale numerical simulations.

One of the most computationally demanding tasks in materials science is electronic structure calculations, which provide fundamental insights into material properties. Density functional theory (DFT) calculations, widely used for simulating electronic states and predicting properties such as band structures and phonon dispersion, involve solving large eigenvalue problems and performing Fast Fourier Transforms (FFT) to compute wavefunctions efficiently \cite{sholl2022density}. These matrix-heavy operations are highly compatible with GPU acceleration. Compared to CPU-based implementations, GPU-accelerated DFT can achieve significant speed increases, allowing researchers to study larger systems or perform high-throughput calculations within a practical time frame \cite{huhn2020gpu,zhang2024gpu}. Beyond first-principles simulations, molecular dynamics (MD) simulations also benefit greatly from GPU acceleration. MD simulations involve integrating Newton’s equations of motion for thousands to millions of atoms, requiring repeated evaluations of interatomic forces, energy minimization, and trajectory propagation over thousands of time steps. The compute-intensive nature of MD simulations makes them an ideal candidate for GPU acceleration, as each atomic interaction can be computed independently in parallel. This level of acceleration has enabled simulations of increasingly complex materials and systems \cite{liu2008accelerating,phillips2020scalable}.

In general, hardware acceleration has become an indispensable tool in modern materials research. It not only enables new AI methodologies that were previously infeasible due to computational constraints but also enhances the efficiency and scalability of high-throughput calculations \cite{deng2024high,chen2016understanding} and large-scale simulations \cite{lu202186,das2023large}. The continued development of specialized AI hardware, such as TPUs and neuromorphic computing architectures \cite{mehonic2020memristors}, alongside heterogeneous computing frameworks that integrate CPUs and GPUs \cite{mittal2015survey}, may allow researchers to train ML models on vast material datasets more efficiently and perform large-scale simulations at lower computational cost in the future. Moving forward, the integration of GPU-accelerated simulations and AI-driven modeling will continue to reshape the way materials are discovered and optimized, making computational materials science an increasingly data-driven, high-performance domain.

\subsection{Datasets}
While necessary and critical, algorithms and hardware alone are insufficient to fully unleash the potential of ML in materials design. High-quality datasets are indispensable for training effective models. Fortunately, data availability has reached a pivotal inflection point where incremental advancements have led to transformative breakthroughs, fueling the ongoing wave of applying ML in materials science. While some challenges in energy materials design are still in the realm of ``small" or ``medium" data problems, the broader trend is moving toward increasingly large and diverse datasets. This growth is propelled by significant advances in high-throughput simulations and experimental techniques, resulting in the development of extensive databases. Prominent examples include but are certainly not limited to the open quantum materials database(OQMD) \cite{saal2013materials},  Materials Project(MP) \cite{jain2013commentary}, phonon databases \cite{petretto2018high}, electronic transport \cite{ricci2017ab}, and TE property databases \cite{gaultois2013data,na2022public,sierepeklis2022thermoelectric}, etc. These meticulously curated data sets are invaluable assets and provide a robust foundation for advancing data-driven methods in the development of energy materials. 

The dataset construction must be tailored to the specific physical problem to ensure more efficient data utilization. For instance, in thermoelectric applications, a dataset that accurately characterizes electrical conduction, heat transport, and the Seebeck effect is critical. Usually, the dataset is constructed from the experimental record. One early attempt from Gaultois \textit{et al.} \cite{gaultois2013data} was the first to carry out a data-driven review, manually compiling the TE properties across various temperatures and families of thermoelectric materials extracted from 100 articles with 1100 data records. Some recent examples include Ref.\cite{na2022public}, which collects 5205 experimental observations covering 880 unique thermoelectric materials with five experimentally measured TE properties: the Seebeck coefficient, electrical conductivity, thermal conductivity, power factor, and $zT$. Recently, AI has also been applied to extract data to construct the targeted datasets. ChemDataExtractor is a chemistry-aware natural language processing toolkit \cite{swain2016chemdataextractor} used for extracting data and conducting data mining from the text. It has been utilized to construct an auto-generated TE dataset by extracting data from the text of 60843 scientific papers containing 22805 data records, which involve the TE figure of merit, the Seebeck coefficient, thermal conductivity, electrical conductivity, and power factor \cite{sierepeklis2022thermoelectric}. Similarly, Springer Materials recently released the database for TE constructed with an AI tool \cite{SpringerMaterials}, which includes thermal conductivity, $zT$, power factor, conductivity, and the Seebeck coefficient. Besides, the LLM-based data extractor also applies to the TE literature \cite{itani2024large} with the GPTArticleExtractor workflow \cite{zhang2024gptarticleextractor} to involve 7123 TE compounds, containing key information such as chemical composition, structural detail, the Seebeck coefficient, electrical and thermal conductivity, power factor, and $zT$. These efforts illustrate that AI can significantly enhance dataset construction through a self-feeding process. 

\definecolor{rulecolor}{RGB}{0,71,171}
\definecolor{tableheadcolor}{gray}{0.92}
%
\definecolor{rulecolor}{RGB}{214,131,7}
\definecolor{tableheadcolor}{gray}{0.92}

\newcommand{\topline}{ %
        \arrayrulecolor{rulecolor}\specialrule{0.1em}{\abovetopsep}{0pt}%
        \arrayrulecolor{tableheadcolor}\specialrule{\belowrulesep}{0pt}{0pt}%
        \arrayrulecolor{rulecolor}}
\newcommand{\midtopline}{ %
        \arrayrulecolor{tableheadcolor}\specialrule{\aboverulesep}{0pt}{0pt}%
        \arrayrulecolor{rulecolor}\specialrule{\lightrulewidth}{0pt}{0pt}%
        \arrayrulecolor{white}\specialrule{\belowrulesep}{0pt}{0pt}%
        \arrayrulecolor{rulecolor}}
\newcommand{\bottomline}{ %
       \addlinespace[-0.8ex]%
      \arrayrulecolor{white}\specialrule{\aboverulesep}{0pt}{0pt}%
        \arrayrulecolor{rulecolor} %
        \specialrule{\heavyrulewidth}{0pt}{\belowbottomsep}}%

\newcommand{\topmidheader}[2]{%
    \addlinespace[0.6ex]
    \multicolumn{#1}{c}{\textsc{#2}}\\%
    \addlinespace[1ex]
}
\newcommand{\midheader}[2]{%
    \addlinespace[-0.6ex]%
        \midrule\topmidheader{#1}{#2}}
\setlist[itemize]{noitemsep, topsep=0pt, partopsep=0pt, leftmargin=1em}
\renewcommand{\arraystretch}{0.8}

\begin{table}[!htbp]
    \centering
    \footnotesize
    \scalebox{1.0}{
    \begin{tabular}[t]{p{4cm} p{3.5cm} p{7cm}} 
        \topline
        \rowcolor{tableheadcolor}
        \textbf{Reference} & \textbf{Num of compounds} & \textbf{TE-related contents \& source}\\
        \midtopline 
        \topmidheader{3}{From literature}
        Gaultois \textit{et al.} \cite{gaultois2013data}  & 1,100 & \vspace{-0.6em} 
        \begin{itemize}
            \item $\sigma, S, \kappa, PF, zT$;
            \item Extracted from 100 research articles.
        \end{itemize}
        \\
        Na \textit{et al.} \cite{na2022public}  & 880 & \vspace{-0.6em} \begin{itemize}
    \item $\sigma, S, \kappa, PF, zT$;
    \item Extracted from 5,205 experimental observations.
\end{itemize}\\
    Sierepeklis \textit{et al.} (ChemDataExtractor) \cite{sierepeklis2022thermoelectric} &  10,641 & \vspace{-0.6em}
    \begin{itemize}
        \item $\sigma, S, \kappa, PF, zT$;
        \item Extracted from 22,805 data records in scientific literature.
    \end{itemize}\\
        SpringerMaterials \cite{SpringerMaterials}  &
        $135\sim$ 2,855 
 &\vspace{-0.6em} 
        \begin{itemize}
        \item $\sigma, S, \kappa, PF, zT$;
        \item Extracted from scientific literature.
    \end{itemize}
    \\
        Itani \textit{et al.} \cite{itani2024large}  & 7,123 & \vspace{-0.6em}  \begin{itemize}
        \item $\sigma, S, \kappa, PF, zT$;
        \item Extracted from $\sim$ 20,000 scientific literature using Large Language Model.
    \end{itemize}         \\ \midheader{3}{From computation}
        Chen \textit{et al.}\cite{chen2016understanding}  & 48,770 &\vspace{-0.6em} \begin{itemize}
        \item $S, \kappa, PF$, band gap ($E_g$);
        \item High-throughput DFT + Clarke \& Cahill-Pohl model calculation.
    \end{itemize}
    \\ 
    Petretto \textit{et al.} \cite{petretto2018high}  & 1,521 &\vspace{-0.6em} \begin{itemize}
        \item Semiconductors; full phonon dispersion, vibrational density of states;
        \item High-throughput Density Functional Perturbation Theory (DFPT) calculation.
    \end{itemize}\\
    Ricci \textit{et al.}
    \cite{ricci2017ab} & 48,000 & \vspace{-0.6em} \begin{itemize}
        \item $\sigma, S, \kappa_{el}$ (electronic contribution only);
        \item DFT + BoltzTraP; Boltzmann transport under constant relaxation time approximation.
        \end{itemize}
        \\ 
    Toher \textit{et al.} \cite{toher2014high}, Elsheikh \textit{et al.} (AFLOWLIB) \cite{elsheikh2014review} & 5,664 & \vspace{-0.6em} \begin{itemize}
        \item $\kappa_{ph}$ (phonon contribution only);
        \item DFT + quasi harmonic approximation (QHA).
    \end{itemize} \\ \vspace{0.6em}\\
        \bottomline
    \end{tabular}}
    \caption{Summary of available datasets related to thermoelectric (TE) quantities.}
    \label{tab1}
\end{table}

High-throughput calculation enables another avenue to obtain a large dataset all from first-principles calculations. For example, Ref.\cite{chen2016understanding} reported thermoelectric properties for more than 48000 inorganic compounds from MP \cite{jain2013commentary}. The electronic structure is calculated through DFT, and TE transport coefficients are calculated using the Boltztrap package \cite{madsen2006boltztrap}. Besides, some specific datasets are also suitable for TE-related studies. These datasets provide partial information for TE, such as first-principles electron transport \cite{ricci2017ab} based on a large set assuming constant relaxation time under the Boltzmann transport theory and a high-throughput density-functional perturbation theory phonons dataset \cite{petretto2018high} with full phonon dispersion and vibrational density of states for 1521 semiconductor compounds in the harmonic approximation. 

However, these TE-related properties are extremely sensitive to defects and microstructure, and further development of datasets for defective structures is necessary. Moreover, enhancing calculation accuracy remains a critical challenge for datasets constructed from high-throughput calculations, and ensuring robust and stable TE materials is crucial for constructing a sustainable TE device for industrial applications. Although comprehensive databases on material stability, mechanical properties, and degradation behavior are essential, these aspects still warrant further exploration with TE.

\subsection{Representation of materials}
Even with sufficient high-quality data, the choice of material representations\textemdash the computer-readable data structures that encode the material information\textemdash can significantly affect data efficiency. This is because different representations encode the physical information of materials in distinct ways. Compared to approaches that rely solely on large datasets or first-principles calculations, integrating physical modeling with data through proper representations is important to reduce the cost of data acquisition and improve the model performance. This is particularly crucial for energy materials research where the computational cost of first-principles calculations can be exceedingly high.

Numerous representations have been developed to describe materials, each tailored to capture different aspects of compositional and structural characteristics \cite{chen2021machine,damewood2023representations}. 
 Among these, the ability to encode defects is critical for accurately modeling defective materials. Although several approaches exist for representing defects, this area remains in its early stages \cite{damewood2023representations}. Here, we discuss a range of representations used to study both general materials and specific defect structures.

\begin{figure}[t]
\centering
\includegraphics[width=0.999\linewidth]{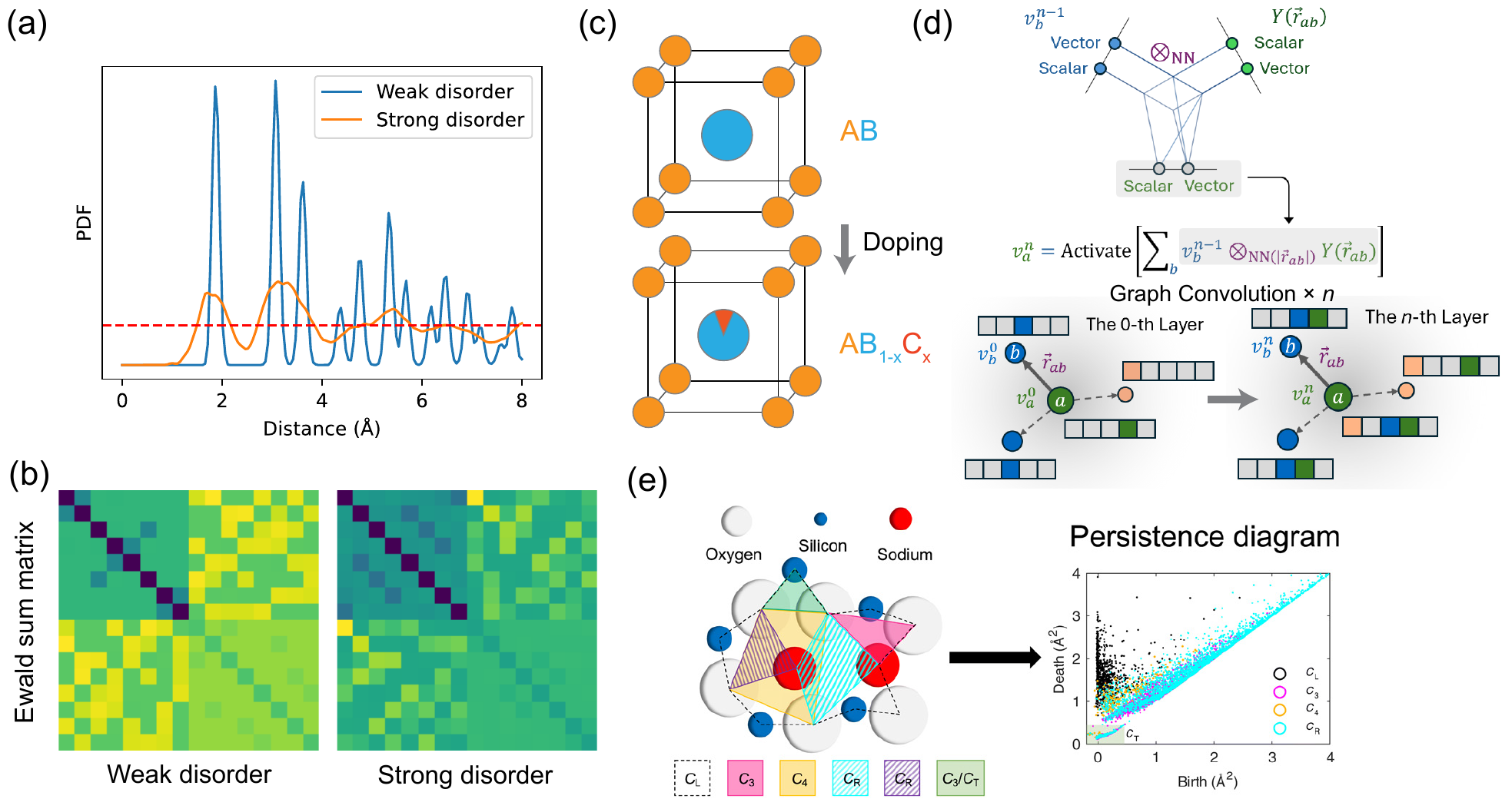}
    \caption{\textbf{Useful descriptors for materials with defect or disorder.} With the Gaussian noise as the local atomic perturbation from the imposed disorder, \textbf{(a)} Pair distribution function (PDF) and \textbf{(b)} Ewald sum matrix presents a notably sensitive response under two different scenarios: weak disorder and strong disorder, introduced through different levels of local atomic perturbation. \textbf{(c)} Sketch for a possible graph representation used in GNN to deal with doping by imposing the doping ratio through the node feature. \textbf{(d)} Illustration of e3nn, an implementation of E(3)-equivariant GNN. Structural information passes within a cutoff radius $r_{\text{max}}$ for a given atom, and the angular information and radial information between two atoms are encoded in spherical harmonics $Y_{lm}(\vec{r}_{ab})$ and a neural network $R(|r_{ab}|)$, respectively. Reproduced with permission of Ref. \cite{chen2021direct}. \textbf{(e)} Persistent homology diagram for amorphous solid. The different colors represent the different dimensional topological features. The location of the data point represents when these topological features appear and disappear with the continuous change of length scale. Reproduced with permission of Ref. \cite{sorensen2020revealing}.} 
    \label{fig:descriptor}
\end{figure}

The pair distribution function (PDF) representation is closely related to data analysis in x-ray and neutron total scattering experiments, where it is used to quantify both short-range and long-range atomic structures \cite{egami2003underneath,zhu2021bridging}. From the experimental side, PDFs can be measured from total scattering, which includes both Bragg diffraction and diffuse scattering to probe local disorder and defects\cite{zhu2021bridging}. The PDF describes the weighted probability distribution of the interatomic distance in real space by providing the likelihood of finding a pair of atoms separated by a distance $r$. The unweighted atomic pair distribution function, $g(r)$, can be defined as:
\begin{equation}
    g(r) = \frac{\rho(r)}{\rho_0} = \frac{1}{4\pi r^2N \rho_0} \sum_{i} \sum_{j \neq i} \delta(r - r_{ij})
\end{equation}
where $\rho(r)$ and $\rho_0$ are the local and average number densities of atoms, respectively, $r_{ij}$ refers to the distance between the $i$th and $j$th atoms in the system with $N$ atoms. A commonly used alternative is the reduced form:
\begin{equation}
    G(r) = 4\pi r^2 \left[\rho- \rho_0\right] = 4\pi r^2 \rho_0 \left[g(r)-1\right]
\end{equation}
An illustrative example is shown in Fig.\,\ref{fig:descriptor}(a), where the peaks of PDF at larger radial distances away from a central atom are suppressed, indicating the absence of long-range order in disordered materials. 
Since the PDF effectively captures defective structures, it has been employed as a key representation in ML studies to investigate defects, such as oxide defects \cite{zhang2023pair}. Besides, the PDF has also been successfully applied in predicting electron and phonon properties in amorphous materials with vacancy defects \cite{cheng2024predicting}.

The Ewald sum matrix generalizes the Coulomb matrix representation in molecules to periodic solids by considering the Coulomb interaction between atoms in a crystal \cite{faber2015crystal}. To account for the Coulomb interaction in a periodic crystal, the Ewald sum matrix introduces a uniform neutralizing background charge \cite{faber2015crystal}, resulting in the following decomposition:
\begin{equation}
    \chi_{ij} = \chi_{ij}^{r} +  \chi_{ij}^{m}+ \chi_{ij}^{0}
\end{equation}
where $\chi_{ij}^{r}$ is the short-range interaction, $\chi_{ij}^{m}$ is the long-range interaction calculated in the reciprocal space, and $\chi_{ij}^{0}$ is a constant correction ensuring the charge neutrality. The short-range interaction term is given by
\begin{equation}
    \chi_{ij}^{r} = Z_i Z_j \sum_{\mathbf{L}}\frac{\mathrm{erfc} ( a | \mathbf{r}_i- \mathbf{r}_j + \mathbf{L}|)}{ | \mathbf{r}_i- \mathbf{r}_j + \mathbf{L}| }
\end{equation}
where $\mathbf{r}_i$ and $Z_i$ are the position of the $i^{\text{th}}$ atom and the atomic number, respectively. $\mathbf{L}$ is the lattice vector, and the sum is taken over all lattice vectors with some real-space cutoff. The long-range interaction part can be written as 
\begin{equation}
    \chi_{ij}^{m} =\frac{ Z_i Z_j }{\pi V}\sum_{\mathbf{G}}\frac{e^{-|\mathbf{G}|^2/(2a)^2}}{ | \mathbf{G} |^2 }\cos\left(\mathbf{G} \cdot(\mathbf{r}_i-\mathbf{r}_j)\right)
\end{equation}
summation is taken over all non-zero reciprocal lattice vectors $\mathbf{G}$ with some reciprocal-space cutoff, while $V$ is the unit cell volume. Currently, the Ewald sum matrix has been used as a structural representation of compositional disorder \cite{yaghoobi2022machine} and has assisted in the theoretical modeling of the displacement fields around point defects \cite{lechner2008displacement}. The features of the Ewald sum matrix are highly sensitive to varying levels of disorder, as illustrated in Fig. \ref{fig:descriptor}(b). In particular, the matrix elements exhibit significant changes as the disorder increases, introduced via local Gaussian noise as a perturbation. This highlights the capability of the Ewald sum matrix to serve as an effective descriptor to capture local ordering variations.
Similarly, another widely-used representation for capturing local atomic structures and defects is the smooth overlap of atomic positions (SOAP) representation \cite{bartok2013representing}. SOAP has proven effective for various ML tasks, particularly in the study of grain boundaries, where it has been extensively applied \cite{fujii2020quantitative}. A detailed discussion of SOAP will follow later in Sec. 4.4 with specific examples.

More recently, the representations of materials have shown a growing preference for graph neural networks (GNNs), which naturally incorporate 3D atomic coordinates and leave huge room for atomic features. 
A graph $G=(V,E)$ consists of a set of nodes $V$ and edges $E$ representing the node connections. More formally, it consists of a set of vertices (or nodes) $v\in V$ and a set of edges $e_{u,v}=(u,v)\in E$, which represent the connections between nodes.
This structural framework provides an intuitive and flexible way to model molecules and materials, where atoms correspond to nodes and chemical bonds form the edges. It is important to note that this does not have to be the case, and the meaning of nodes can be broader than atoms even in materials science \cite{okabe2024virtual}. The fundamental idea behind GNNs is that atomic interactions within local neighborhoods influence the overall properties of a material. 
GNNs are trained using Message-Passing Neural Networks (MPNNs), which iteratively aggregate and exchange information across nodes and edges \cite{gilmer2017neural}. Due to their adaptability and efficiency, GNNs have become a powerful tool for predicting material properties \cite{reiser2022graph}.

Despite the success of early GNNs such as Crystal Graph GNN (CGCNN) \cite{xie2018crystal} propagating information through graph convolutions, there is still room for improvement in terms of symmetry; na\"ive GNNs only guarantee atomic permutation symmetry for global property prediction. A more robust representation of materials in 3D space is to preserve the 3D Euclidean group E(3) equivariance, including translation, rotation, and inversion. Formally, a function $f:\ X\rightarrow Y$ is equivariant with respect to a group $G$ (in this case, the E(3) group) that acts on $X$ and $Y$ if:
\begin{equation}
D_Y[g]f(x)=f(D_X[g] x), \forall g \in G, \forall x\in X    
\end{equation}
where $D_X, D_Y$ are the representations of group $G$ acting on $X$ and $Y$, respectively.

The power of equivariant GNNs lies in their capability to incorporate equivariant operations to preserve the known transformation properties of physical systems under coordinate changes \cite{satorras2021n}.
These properties include scalars (e.g., energy and band gap), vectors (e.g., force and polarization), and rank-2 tensors (e.g., dielectric and conductivity tensors). By inherently respecting these symmetries, equivariant GNNs can enhance data efficiency by orders of magnitude without the need for manual data augmentation.
There are numerous strategies implementing GNNs with E(3) equivariance, and the realization of e3nn is briefly introduced here. The core idea of e3nn is to represent features of various orders (scalars, vectors, rank-2 tensors, etc.) using spherical harmonics, which are inherently equivariant under E(3) operation. 
To preserve E(3) equivariance in e3nn is to separate input encodings into radial and angular components and propagate the nodal information via tensor product \cite{geiger2022e3nn}, i.e.
\begin{equation}\label{e3nn}
    v'_i =\frac{1}{\sqrt z}\sum_{j\in\partial(i)}{v_j\otimes\ \text{NN}(||{\vec{r}}_{ij}||)Y({\vec{r}}_{ij}/||{\vec{r}}_{ij}||)}
\end{equation}
where $v_i$ and $v'_i$ are node features before and after the graph convolution layer of atom $i$, respectively, and the summation is over all neighboring atoms connected within a cut off radius ($r_{\max}$), $\vec{r}_{ij}$ is the edge vector pointing from $i$ to $j$, NN is a parametrized neural network encoding the radial distance information and $Y$ is the spherical harmonics encoding the angular information (Fig.\,\ref{fig:descriptor}(d)).
During the message passing step in Eq.\,\ref{e3nn}, e3nn decomposes the tensor product of spherical harmonics into a direct sum of spherical harmonics of different orders according to the Clebsch-Gordan formula. This ensures E(3) equivariance throughout every layer in e3nn.
Equivariant GNNs like e3nn have demonstrated great efficiency and accuracy in predicting materials properties, such as phonon density-of-states \cite{chen2021machine}, optical response \cite{hung2016quantum}, and electron density-of-states \cite{kong2022density}. 

The development of defect representation also benefits from the advancement of GNNs. 
Intuitively, point defects, such as dopants, can be encoded by mixing the original node embedding with the dopant atom embedding, as shown in Fig.\,\ref{fig:descriptor}(c). Recent studies have successfully applied GNNs for defect prediction in semiconductors \cite{rahman2024accelerating,xiang2024exploration}. Instead of representing a crystal structure as a point cloud of atoms, one can also view it as a point cloud of defects. This sparse representation with GNN provides a novel way to describe and model defective structures. In fact, this approach has been used to efficiently and accurately predict the performance of 2D materials \cite{kazeev2023sparse}. 

Usually, robust information is embedded in the topology of the atomic structure. The topological descriptor is introduced to reveal this hidden information, such as several topological data analysis techniques.
Recently, one of the topological data analysis techniques, persistent homology or its duality, persistent cohomology \cite{edelsbrunner2008persistent}, has shown great potential to represent complex structures with defects as a topological descriptor.
Features of persistent homology are generated from the topological invariants of the homology group within different dimensions. This topological invariant can be understood as the number of holes or voids within the corresponding dimensions, such as a circle having one hole in 1D, while a spherical surface has one hole in 2D. As a result, the topological invariants in different dimensions comprehensively depict the topological feature. On the other hand, by adjusting the length scales to determine the size of local cover generated by each data point, the union of all local covers collectively defines the different manifolds. Since the homology group captures the corresponding topological feature of the manifolds, 
the range of length scales can describe how this feature of the different dimensional holes or voids evolves while they continuously appear (persistences) \cite{damewood2023representations}.
Results from persistent homology are often displayed as persistence diagrams, as is shown in Fig.\,\ref{fig:descriptor}(e). The persistent diagram provides global information on the topological manifold of the data, which is robust to external perturbation. 
This makes persistent homology a powerful tool for uncovering hidden patterns in amorphous solids, where the MD trajectory was utilized to construct a persistent diagram and recover the hidden medium-range order \cite{sorensen2020revealing}.
Besides, incorporating the GNN with persistent homology features could significantly increase the prediction accuracy of the defect-related property, such as the defect formation energy \cite{fang2025leveraging}. 
It is also worth noting that persistent cohomology has been instrumental in detecting disorder through the interpretation of spectroscopic data. For example, it has been used to identify disorder-induced obstructions in Majorana zero modes in nanowires by scanning tunneling spectroscopy \cite{cheng2024machine} and to classify bacterial strains based on Raman spectroscopy \cite{offroy2016topological}.

\section{Thermoelectric Design with Defects and Machine Learning}

The increasing power of AI is transforming the optimization process, particularly in high-dimensional optimization problems. In bulk TE material design, ML enables more efficient exploration of complex defect landscapes, helping identify optimal configurations. This section categorizes ML-driven strategies for bulk TE design based on different aspects of defect engineering.

Defect engineering is one of the most important approaches to improve TE materials \cite{zheng2021defect,wu2024defect,beretta2024strategies,han2024advancements}. Controlling defects allows the structure and properties of the materials to be tuned to optimize performance. However, defects introduce structural complexities and destroy the perfect structure, making it challenging to model the precise relationship between material performance and defect tunability. As a result, while defect engineering in bulk TE materials has gained increasing attention, much of the progress remains experimental, and the theoretical or computational guidance is relatively limited. 

The emergence of ML brings new opportunities. Although defects introduce extra complexity and degrees of freedom to materials, ML excels at modeling high-dimensional functions with complex relations. There are two critical components to successfully apply ML to defect engineering: high-quality data and effective representations for defective structures \cite{damewood2023representations}, which will be discussed in the following subsections.

\subsection{Design point defects for thermoelectric performance}
\begin{figure}[!htbp]
    \centering
    \includegraphics[width=0.95\linewidth]{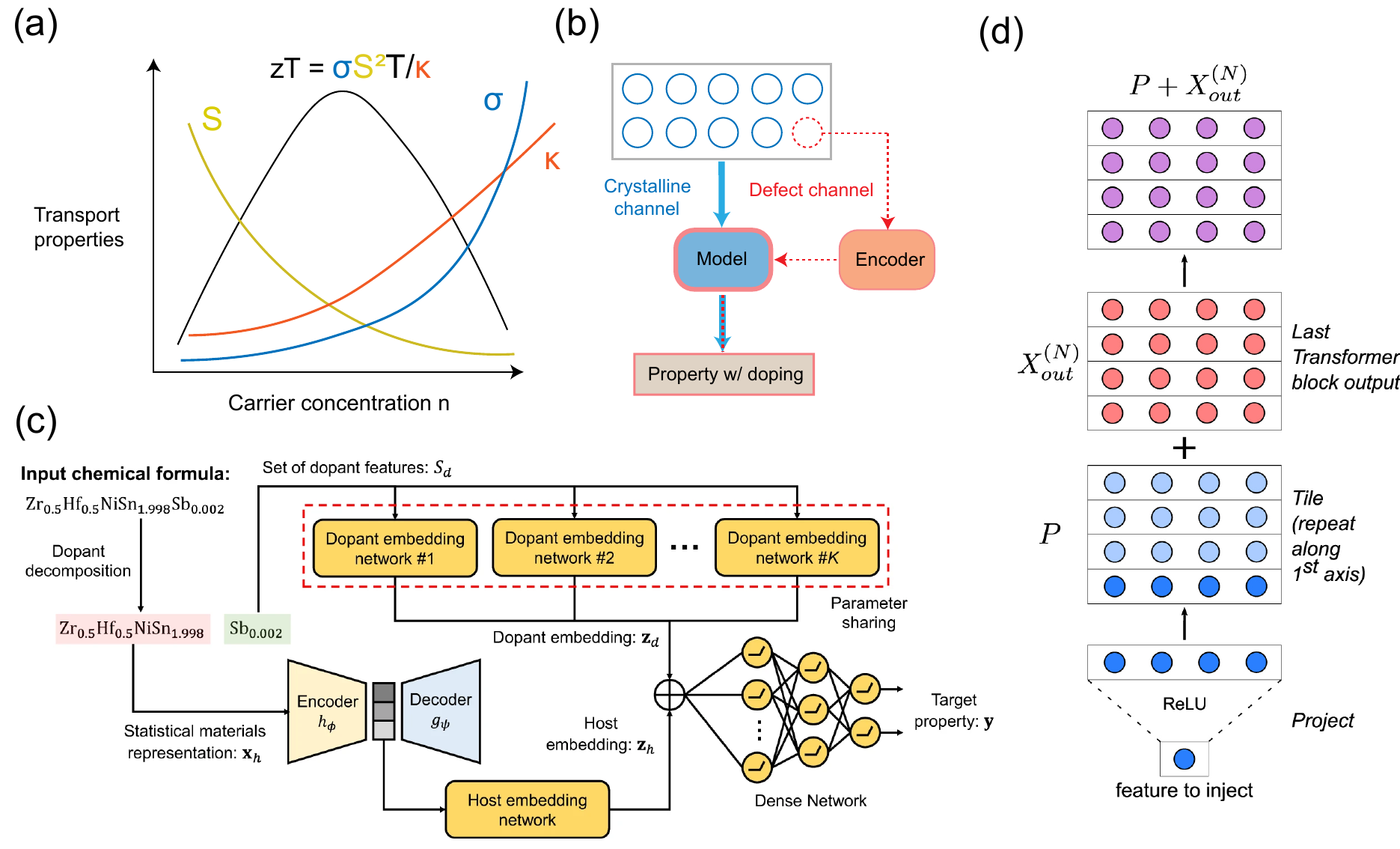}
    \caption{
    \textbf{Machine learning (ML) aided defect (doping) engineering for TE.} \textbf{(a)} Nonlinear behavior of TE-related transport properties w.r.t. the doped carrier concentration; \textbf{(b)} Schematic illustration of ML strategy to emphasize defect information by introducing a separate defect channel with nonlinearity; \textbf{(c)} The ML network architecture of DopNet, passing on dopant and host embedding separately before merging them to a dense network. Reproduced with permission of Ref.\cite{na2021predicting}; \textbf{(d)} The key ML architecture of CraTENet, improving the expressibility of defect channel using the transformer blocks. Reproduced with permission of Ref.\cite{antunes2023predicting}.}
    \label{fig:defect}
\end{figure}

The zero-dimensional point defect is a fundamental type of defect in materials.
Point defects are localized disruptions in the periodic atomic arrangement of crystalline solids, including vacancies, interstitials, and substitutional defects. 
Even in the lowest-dimensional case, the combination of these point defects rapidly increases the complexity of defect engineering.
Point defects can be classified as intrinsic (formed naturally, e.g., vacancies) or extrinsic (induced deliberately in labs, e.g., doping). 
Despite their small scale, point defects play a pivotal role in tuning TE materials. Doping, in particular, has emerged as the most common strategy for optimizing the $zT$ of a TE material \cite{ma2021review,wei2020review,xiong2024doping}, and advancements in first-principles calculations now provide a framework to predict the doping effects and guide the doping strategy \cite{gonzalez2018variation,qu2020doping,li2022doping}.
However, the relationship between doping and TE performance is complex. While doping can enhance the carrier concentration $n$ and thereby increase electrical conductivity $\sigma$, it suppresses the Seebeck coefficient $S$ due to the reduced entropy effect. From Eq. \ref{zT}, there exists an optimal doping level to maximize $zT$ (Fig. \ref{fig:defect}(a)).
Empirically, the doping level for ideal TE properties is within $10^{19}\sim 10^{21} \rm{cm}^{-3}$\cite{snyder2008complex}, although in certain topological materials the optimal carrier concentration could be much lower, down to $10^{17} \rm{cm}^{-3}$ \cite{skinner2018large,han2020quantized}. Ref. \cite{beretta2024strategies} offers a brief review of the doping effects in TE materials.

The engineering of point defects often requires advanced computational techniques to first model their structures. Before discussing ML-driven modeling for $zT$, this foundational step is briefly outlined here.
For dopants, random substitution to construct a dataset is not a viable strategy because it can lead to chemically or structurally implausible configurations. Factors such as atomic size mismatch, charge balance, and site preference are often neglected, resulting in unrealistic material compositions.
Even selecting the suitable elements for substitution is non-trivial. 
According to the Goldschmidt rules of substitution \cite{goldschmidt1926gesetze}, ions with similar radii and charges are the most favored candidates for substitutional doping.
Traditionally, the compatibility between the host and the dopant ions is often evaluated using the similarity of their Shannon ionic radii, which depend on their oxidation state and coordination environment. The ionic radii database, originally compiled by Shannon \cite{shannon1969effective,shannon1976revised}, has been expanded from 475 ions to 987 ions using ML techniques \cite{baloch2021extending}.
However, this approach can yield unreasonable predictions due to its oversimplification of interatomic interactions. Additionally, there is a lack of a quantitative measure governing dopant selection, as the probability of sampling dopants is not explicitly well-defined within this qualitative framework.
To address this challenge, Hautier \textit{et al.} proposed an ML-based probabilistic model to enhance the discovery of new compounds through ionic substitutions, focusing on retaining the crystal structure of the original material \cite{hautier2011data}. 
This method extends the traditional concept of ionic similarity by formulating substitution likelihoods as a probabilistic function. 
Trained on an extensive database of crystal structures, this method uses a probabilistic model to evaluate the compatibility of ions by tokenizing them as a machine translation task.
The framework outperforms heuristic methods like Goldschmidt's rules by offering quantitative metrics into substitution tendencies for doping. It should also be noted that both the substitional probability method and the Shannon radii similarity formalism are implemented in the pymatgen library\cite{ong2013python}.

Although there has been significant progress in the application of ML directly to predict doping-enhanced TE performance from the literature \cite{parse2022machine}, a generic predictive model may still begin with the \textit{ab initio} method. Once suitable dopants are identified, atomistic simulations such as DFT can be employed to model the structural and electronic properties of the doped material, making it possible for high-throughput screening and ML property prediction.
However, while GNNs have shown good performance in predicting the properties of crystalline materials \cite{reiser2022graph}, ML still faces several challenges when applied to doped materials.
One major challenge is the modeling of the defect structure\textemdash the supercell approach is intrinsically periodic and does not accurately represent the random distribution of defects. It cannot represent irrational defect concentrations. Moreover, at low defect concentrations, the supercell approach has a high computational cost. For complex TE materials in particular, properties like $zT$ can be extremely sensitive to the doping level; even a small variation can drastically affect the material's performance \cite{irfan2024advancements,pei2012band}.
Moreover, such changes are often nonlinear with respect to defect concentration, making it difficult for traditional ML models to predict the associated effects with high fidelity, especially when minor variations in the input representation of materials result in significantly altered properties.

Over the past few years, advancements in ML have addressed challenges in modeling the nonlinear impact of dopants on TE properties, by explicitly incorporating defect information into the input data (Fig. \ref{fig:defect}(b)).
Na \textit{et al.} introduced DopNet, a neural network designed to predict doped TE materials \cite{na2021predicting}. The model separately embeds the host material and the dopants, using an architecture composed of host embedding networks, dopant embedding networks, and a dense prediction layer (Fig. \ref{fig:defect}(c)). Such an approach ensures an explicit representation of dopant effects and addresses the challenge of vanishing dopant effects in traditional representations. 
DopNet outperforms other ML models like gradient-boosted regression trees (GBRT) and Gaussian process regression (GPR), achieving a mean absolute error (MAE) of 0.06 for predicting $zT$.
Since then, several other approaches have been made targeting specific systems to improve TE properties through ML-driven methods:
A deep neural network potential (NNP) approach was proposed for predicting the lattice thermal conductivity of PbTe systems with intrinsic point defects, providing insights into defect-strain interactions and their contributions to TE performance \cite{qin2023machine}; A GBRT model was developed to predict $zT$ for doped BiCuSeO systems. The model effectively identifies high-performance doping strategies, including Po- and Cs-doped BiCuSeO, achieving $zT$ enhancements by up to 104\% compared to undoped counterparts \cite{he2023prediction}; Key TE properties, including Seebeck coefficient, electrical conductivity, and $zT$, are predicted using an extreme gradient boosting regressor (XGBR) and feature engineering on structural representations. The model achieved high accuracy across various types of materials, including chalcogenides and skutterudites \cite{minhas2024machine}.
One advanced ML architecture used for defect engineering of TE was proposed by Antunes \textit{et al.} (Fig. \ref{fig:defect}(d)), where the Compositionally-restricted attention-based ThermoElectrically-oriented Network (CraTENet) was designed to scan a large inorganic materials space for novel TE materials design \cite{antunes2023predicting}. The transformer architecture allows CraTENet to capture intricate relationships between elemental compositions and TE properties by leveraging self-attention mechanisms, which prioritize the most relevant features of the input data \cite{vaswani2017attention}.
By training on high-throughput DFT data, the model predicts TE properties across diverse material systems with an MAE as low as 49 $\mu V \cdot K^{-1}$ for the Seebeck coefficient.

Despite the advancement of ML in doping engineering for TE, current ML-aided workflows still face a few challenges. 
One challenge is the reliance on composition-only input representations, which may overlook critical structural information, particularly in systems where local atomic configurations or defect-site geometries play a crucial role. 
Although direct implementation of GNNs is inefficient for modeling defective structures, there are some promising strategies to include defect information in graph-based architectures.
One approach is to introduce defect information as a conditional channel in the message-passing layers of the GNN. Alternatively, virtual nodes could be incorporated into GNNs to make the model aware of defect geometry \cite{gilmer2017neural,pham2017graph}. The use of virtual nodes has proven their expressibility power in full phonon bandstructure prediction \cite{okabe2024virtual}. 
Another key challenge is the limited availability of data on defective TE materials, which will be discussed in the ``Outlook" section.

\subsection{Guide entropy engineering for thermoelectric design}
Motivated by the development of high-entropy alloys \cite{miracle2017critical,george2019high}, recent advances in compositional complexity with high configurational entropy have been applied to enhance functional properties such as TE \cite{jiang2023high,ghosh2024high}, battery \cite{ouyang2024rise}, hydrogen storage \cite{yang2022recent}, among others. Specifically, entropy engineering has gained increasing attention as a powerful strategy for optimizing TE materials \cite{jiang2023high,ghosh2024high,moshwan2024entropy,tang2024high,liu2025high}. Several experimental results have reported superior TE performance in materials tuned by entropy engineering \cite{qiu2019realizing,jiang2021high,jiang2021entropy,jiang2022high,ma2022high}. The underlying mechanisms can be summarized as follows: the high-entropy materials may form a large lattice mismatch due to their multi-element composition, which enhances phonon scattering and effectively reduces lattice thermal conductivity. Simultaneously, the large compositional space enables precise tuning of symmetry and electronic band structures to optimize electrical transport properties\cite{liu2017entropy,moshwan2024entropy}. However, phase boundaries in multiphase regions can serve as scattering centers that hinder charge carrier transport, negatively impacting electrical conductivity. This highlights the importance of achieving a stable single-phase region in high-entropy thermoelectric materials, as single-phase stabilization helps to maintain high electrical conductivity while still benefiting from reduced thermal conductivity, thereby enhancing the overall figure of merit \cite{ghosh2024high}. Besides, the high-entropy phase also provides a platform to tune the microstructure through the multiple different local chemical orderings. All of these factors play an important role in shaping the TE performance. However, navigating the vast compositional design space of high-entropy materials remains a formidable computational challenge. Even determining the phase stability of high-entropy materials within this vast compositional landscape is a nontrivial task. 

The CALPHAD (CALculation of PHase Diagrams) method is a state-of-the-art thermodynamic modeling approach that can predict high-entropy phase stability from experimental and first-principles data \cite{lukas2007computational,zhang2022calphad}. Recent refinements, such as incorporating chemical short-range order \cite{sundman2018review,fu2024cluster}, further improve its accuracy in phase predictions by combining more physical thermodynamic modeling. The ability of CALPHAD to handle multicomponent thermodynamics of the high-entropy materials allows it to predict high-entropy phase diagrams and their thermodynamic functions to provide valuable insights into optimizing doping strategies, entropy engineering, and microstructures, ultimately guiding the optimization of TE performance.
As a result, CALPHAD has recently been considered and utilized as one of the most important design toolkits for high-entropy TE materials design \cite{li2021calphad,ghosh2024high}.

\begin{figure}[t]
    \centering
\includegraphics[width=0.95\linewidth]{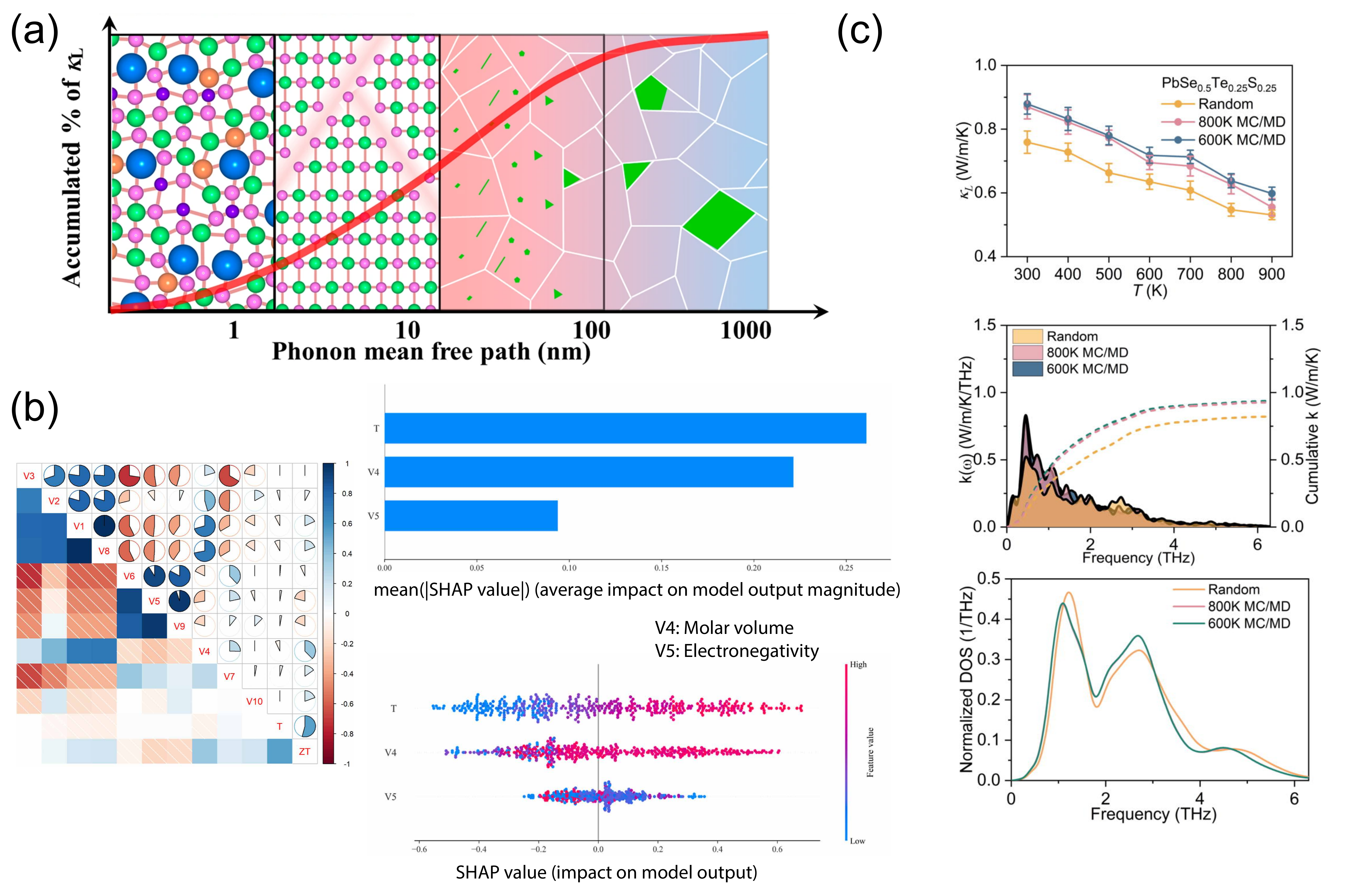}
    \caption{\textbf{Entropy engineering for TE materials.} \textbf{(a)} Phonon mean free path dependence of accumulated lattice thermal conductivity (the red line) and schematic diagram of all-scale hierarchical structures from the high-entropy matrix. It schematizes the ability to improve TE performance in all-scale hierarchical structures through entropy engineering, with permission of reproduction from \cite{jiang2021entropy}. \textbf{(b)} Explainable machine learning for entropy engineering. The left visualization is a Pearson correlation map for features, where blue and red colors indicate positive and negative correlations. The average impact on model output magnitude and the effect of each feature on the output of the model is presented on the right through the SHAP values, with permission of reproduction from \cite{li2023interpretable}. \textbf{(c)} The machine learning (ML) enabled atomic simulation for different local chemical ordering. From top to bottom, thermal conductivities at elevated temperature, spectral thermal conductivity and the cumulative thermal conductively, normalized phonon density of state (DOS) at 300K are presented for random, 800, and 600 K annealed PbSe$_{0.5}$Te$_{0.25}$S$_{0.25}$, with permission of reproduction from \cite{lyu2024effects}.}
    \label{fig:entropy}
\end{figure}

However, CALPHAD is inherently limited in capturing functional properties essential for TE performance, such as electronic band structure and phonon transport \cite{li2021calphad}. Additionally, entropy engineering is usually combined with the hierarchical defect structure to enhance the TE performance, making the simulation even more difficult \cite{jiang2021entropy}, as shown in Fig.\,\ref{fig:entropy}(a). The complexity of these relationships, combined with the intricate effects of local atomic arrangements and disorder, poses a significant challenge on guiding and optimizing functional performance. To tackle this challenge, one of the most promising directions is the application of modern ML algorithms, which go beyond CALPHAD’s data-driven framework by directly enabling simulation and predictions of the transport properties from the related data. In this sense, CALPHAD can be considered an early form of ML-assisted modeling, as it integrates physical principles with data-driven analysis only for multicomponent thermodynamics. However, ML-driven approaches take a step further by learning complex structure-property relationships beyond what CALPHAD can explicitly model, providing deeper insights into TE material performance and enabling the accelerated discovery of high-efficiency TE materials.

Combining first-principles calculation and ML method, people can construct the DFT workflow targeting entropy-engineered TE materials with the machine-learned force field such as Ref.\,\cite{xia2022investigation}. In this work, DFT-based special quasirandom structure (SQS) method \cite{zunger1990special} was used to simulate high-entropy disordered configurations, in combination with on-the-fly ML \cite{podryabinkin2017active,jinnouchi2020descriptors}. Since entropy engineering is mostly employed to optimize thermal conductivity for TE performance, predicting lattice thermal conductivity could benefit TE materials optimization. Therefore, the machine-learned force-field \cite{verdi2021thermal} can be constructed on-the-fly and used to study thermal transport in high-entropy materials to search for high $zT$ candidates. Similarly, it can be integrated between non-equilibrium MD and ML to predict the lattice thermal conductivity of high-entropy materials \cite{lu2024estimating} with the composition of each element and the length of the model after relaxation as the features. However, in this work, ML is not used to construct the potential; instead, it leverages data from MD simulations to directly predict thermal conductivity as a function of composition. Moreover, it is important to note that thermal conductivity alone does not determine TE performance, and electrical transport properties must also be considered. Recently, several attempts have been made to explore the electronic structure and electronic transport in high-entropy materials \cite{mu2018electronic,mu2019influence,mu2019uncovering} utilizing the SQS method and Korringa-Kohn-Rostoker (KKR) Green’s function method combined with the coherent potential approximation (CPA) \cite{stocks1978complete}.  In the future, more emphasis should be placed on assessing entropy-engineered electronic structures for TE application, which is important but has not received enough attention so far. 

The literature mentioned above is mainly on combining DFT or MD with ML to predict the TE performance of high-entropy materials. While experimental data is relatively limited, it is still possible to apply ML to study entropy engineering and TE. Recently, Li and Liu \cite{li2023interpretable} developed an interpretable ML framework to evaluate and optimize the TE performance of high-entropy GeTe-based materials based on an experimental dataset \cite{na2022public}. Feature selection techniques, such as Pearson correlation and univariate feature selection, were applied to represent the most important features, including temperature, the average molar volume, and the average electronegativity. By combining with the SHAP (SHapley Additive exPlanations) method \cite{10.5555/3295222.3295230}, the study identifies key features influencing the TE performance, such as the average molar volume and electronegativity, and provides insights into their impact across different compositional configurations.  This approach enables systematic exploration of high-entropy regions within the relevant phase diagram, facilitating the search for compositions with enhanced TE performance, as shown in Fig.\,\ref{fig:entropy}(b). While this study primarily employs averaged atomic properties as descriptors, future extensions could incorporate local structural features to further refine the predictions. The successful application of ML in this context highlights the potential of entropy engineering as a powerful strategy for TE material optimization, even when experimental datasets are limited.

Entropy engineering not only introduces large configurational entropy to tune the phonon scattering, thermal transport, and electrical properties but also the local chemical ordering, which is yet overlooked. Such a tiny but always-existing effect usually requires detailed study to quantify, such as ML-based simulation. With ML interatomic potential, people recently conducted hybrid Monte Carlo and MD simulations to observe how local chemical ordering interacts with phonon scattering \cite{lyu2024effects}. The thermal conductivities and the related phonon density of the state are calculated by ML-enabled atomic simulation for the fully random microstructure and the quenched microstructures obtained at different temperatures, presented in Fig.,\ref{fig:entropy}(c). The evidence reveals that local chemical order can reduce phonon scattering in the frequency range of 0–2THz, thus enhancing thermal conductivity by approximately 14\%. This observation is achieved efficiently through the efficient ML neuroevolution potential (NEP) for the Pb–Se–Te–S system based on DFT calculations \cite{fan2021neuroevolution}. This ML potential uses an evolutionary strategy to perform large-scale MD simulations and is specialized in heat transport in materials with strong phonon anharmonicity or spatial disorder. The training dataset for the ML interatomic potential includes binary, ternary, and quaternary configurations. The graphics processing unit molecular dynamics (GPUMD) package was adopted to construct the neuroevolution potential efficiently \cite{fan2017efficient,fan2021neuroevolution,fan2022gpumd}. However, this work only presents the comparison between the annealed samples from 600K and 900K to illustrate the impact of local chemical order. The detailed quantitative relation between local chemical order and phonon scattering may require further study through efficient simulation and proper description of the local chemical order, such as the cluster probability distribution proposed in thermodynamic modeling \cite{kadkhodaei2021cluster,fu2024cluster}. Besides, the role of local chemical order in tuning the electronic band structure and its subsequent influence on TE performance remains subtle and is still awaiting further investigation.

 In summary, as more datasets for high-entropy TE materials continue to be generated, the size of the dataset poses great challenges on the entropy engineering for TE materials due to the huge design space from the compositional complexity. ML work based on limited experimental data \cite{li2023interpretable,yadav2022interval} has been conducted.
On the other hand, ML can be used as a surrogate model to construct ML potentials and force fields, enabling more efficient simulations to study transport properties and determine TE performance \cite{xia2022investigation,verdi2021thermal,lyu2024effects,lin2024composition}. A similar strategy involves applying ML to establish the relationship between composition and transport properties using data collected from MD simulations, as mentioned \cite{lu2024estimating}. However, these methods remain constrained by the specific combinations of elements used in the simulations due to the high computational demand. Recently, foundational ML models have shown significant progress, particularly in the development of universal atomic potentials \cite{chen2022universal,takamoto2022towards,choudhary2023unified,batatia2023foundation,zhang2024pretraining,zhang2024dpa}. One can envision that such foundational models covering all common elements from the periodic table will significantly accelerate the search for high-entropy TE materials in the future.

From the perspective of ML algorithms, current exploration remains largely confined to traditional ML and simple deep neural networks, with limited specialization for entropy engineering, apart from applications in ML potentials. These interatomic potentials are generally independent of the specific advancements in ML tailored for high-entropy materials. While previous literature has highlighted challenges in applying GNNs to high-entropy materials—possibly due to the extra interactions caused by chemical short-range order \cite{liu2023machine}—recent efforts have begun to address these issues. Researchers are now exploring the use of GNNs to study high-entropy materials by incorporating additional structural descriptors to capture the chemical short-range order \cite{zhang2024graph}.

\subsection{Dislocations in thermoelectric design}

Dislocations are 1D line defects in crystalline materials. Many experimental efforts have been made to enhance TE performance through dislocation engineering \cite{wu2020thermoelectric,abdellaoui2021parallel,xu2022dense}. Compared to point defects and entropy engineering, dislocations are much more challenging to model for ML due to the lack of efficient structure representations and the lack of high-quality data.

While point defects and entropy engineering can leverage data from high-throughput DFT calculations, the study of dislocations usually requires simulations in significantly larger super-cells, posing challenges for DFT calculations \cite{wang2014polymorphism}. Moreover, representations for point defects or high-entropy materials primarily rely on the composition or other atomic representations; yet in contrast, dislocations, as line defects, present a far greater challenge in developing appropriate and effective descriptors due to their extended nature and the topological constraint. Dislocation density can certainly serve as one possible representation, although many other features of the dislocations will be neglected. Due to the importance of dislocation in mechanical properties, some attempts of embedding dislocation for ML come from mechanical and metallurgy research to study plasticity, such as the work by Salmenjoki \textit{et al.} \cite{salmenjoki2018machine}. In this work, multiple dislocations are embedded for ML through the density of geometrically necessary dislocations with positive and negative Burgers vectors and take its Fourier transform coefficient as the feature (under the periodic boundary conditions). The extensive database of stress-strain curves and the corresponding initial dislocation configurations is generated from the 2D discrete dislocation dynamics (DDD) simulations. With the help of this successful embedding, this work explores the predictions of the correlation between the dislocation configurations and the stress-strain curves, as shown in Fig.\,\ref{fig:boundary}(a). Through this approach, the study explores the predictability of stress-strain responses based on dislocation arrangements, using data from DDD simulations. The findings suggest that ML can effectively capture the interplay between microstructural features and macroscopic mechanical properties, offering a pathway for future investigations into functional properties and optimization of materials such as $zT$.

While high-quality datasets on dislocations remain limited, adopting a similar ML-aided multi-scale simulation approach could drive significant progress in the coming years. A recent study has employed ML to learn dislocation dynamics from MD simulations \cite{bertin2024learning}. In this work, an ML-based surrogate model is trained to connect the dislocation dynamics simulation and MD simulation to benefit the larger-scale modeling of dislocations. For example, thermal transport can be studied through MD simulation \cite{sun2020molecular} and interatomic potential-based larger-scale simulation for the long-range nature associated with the phenomenon of interaction of phonons with fast-moving dislocations \cite{chen2017effects}. Both are possible to be generalized with the ML-assisted larger-scale simulations in many systems to accelerate the TE materials design.

\subsection{Grain boundaries and interfaces}
Grain boundaries and interfaces are 2D planar defects within the 3D materials system. Many attempts have been made to improve TE performance through grain boundary engineering or interface engineering \cite{nandihalli2023imprints,kim2015dense,lin2020expression,tian2021enhanced,rahman2024grain}. However, similar to the case of dislocation, predictive design of TE materials with theory or computation is still challenging. While low-angle grain boundaries can be considered as arrays of dislocations, it is also expected that grain boundaries and interfaces pose similar challenges as dislocations when applying ML techniques, particularly in terms of representation construction and data preparation.

Recently, there has been notable progress in developing representations for grain boundaries \cite{fujii2020quantitative}. The work proposes using the SOAP descriptor to construct a microscopic structure metric, the local distortion factor (LDF). This metric quantifies deviations in the local atomic environment of an atom near a grain boundary compared to that of an equivalent atom in the bulk crystal. SOAP, mentioned earlier as one of the descriptors, encodes local atomic environments by expanding atomic densities in terms of radial and spherical basis functions, providing a rotationally and translationally invariant fingerprint of atomic configurations \cite{bartok2013representing}. Mathematically, the atomic density around a central atom is represented as a sum of Gaussian functions centered on its neighboring atoms, which is suitable for describing defects such as grain boundaries. This density is then expanded using orthonormal radial basis functions and spherical harmonics, yielding a set of expansion coefficients. With a basis of radial functions $g_n (r)$ and spherical harmonics $Y_{lm} (\theta, \phi)$, $\rho(\mathbf{r})$ for central atom $i$ then can be expressed as
\begin{equation}
\rho_{i}(\mathbf{r}) = \sum_{j} e^{-\frac{|\mathbf{r}-\mathbf{r}_{ij}|^2}{2\sigma^2}} = \sum_{nlm} c_{nlm} g_{n}(r) Y_{lm}(\theta,\phi),
\end{equation}
$c_{nlm}$ is the expansion coefficient in the density function. The SOAP then can be defined as the overlap of the two local atomic neighbor densities, integrated over all three-dimensional rotations $\hat{R}$:
\begin{equation}
    \tilde{K}(\rho_i(\mathbf{r}),\rho_k(\mathbf{r})) = \int d \hat{R}\left|\int\rho_i(\mathbf{r})\rho_k(\hat{R}\mathbf{r})d\mathbf{r}\right|^2
\end{equation}
For most applications it is
helpful to normalize this function so that the self-similarity of any environment is unity, giving the final function
\begin{equation}
    K(\rho_i(\mathbf{r}),\rho_k(\mathbf{r}))= \frac{\tilde{K}(\rho_i(\mathbf{r}),\rho_k(\mathbf{r})) }{ \sqrt{\tilde{K}(\rho_i(\mathbf{r}),\rho_i(\mathbf{r}))\tilde{K}(\rho_k(\mathbf{r}),\rho_k(\mathbf{r}))}}
\end{equation}
Finally, it can be computed and simplified for comparing environments based on the local atomic density as \cite{de2016comparing,bartok2013representing}:
\begin{equation}
\begin{split}
    K(\rho_i,\rho_k) &= \hat{\mathbf{p}}_i(\mathbf{r})\cdot \hat{\mathbf{p}}_k(\mathbf{r}),\\
    p_i(\mathbf{r})_{n n' l} &= \pi \sqrt{\frac{8}{2l+1}}\sum_{m}(c_{nlm})^{\dagger}c_{n'lm},
\end{split}
\end{equation}
where $\hat{\mathbf{p}}_i(\mathbf{r})$ represents the unit vector of the local environment collected by the elements of the power spectrum by $p_i(\mathbf{r})_{n n' l}$, commonly referred to as the power spectrum. It serves as the final descriptor, ensuring rotational invariance. With its ability to capture subtle differences in atomic arrangements, SOAP exhibits remarkable versatility in materials applications, making it a powerful tool for tasks such as structure-property predictions, interatomic potential development, and (defected) atomic environment embedding.

\begin{figure}[t]
    \centering
\includegraphics[width=0.9\linewidth]{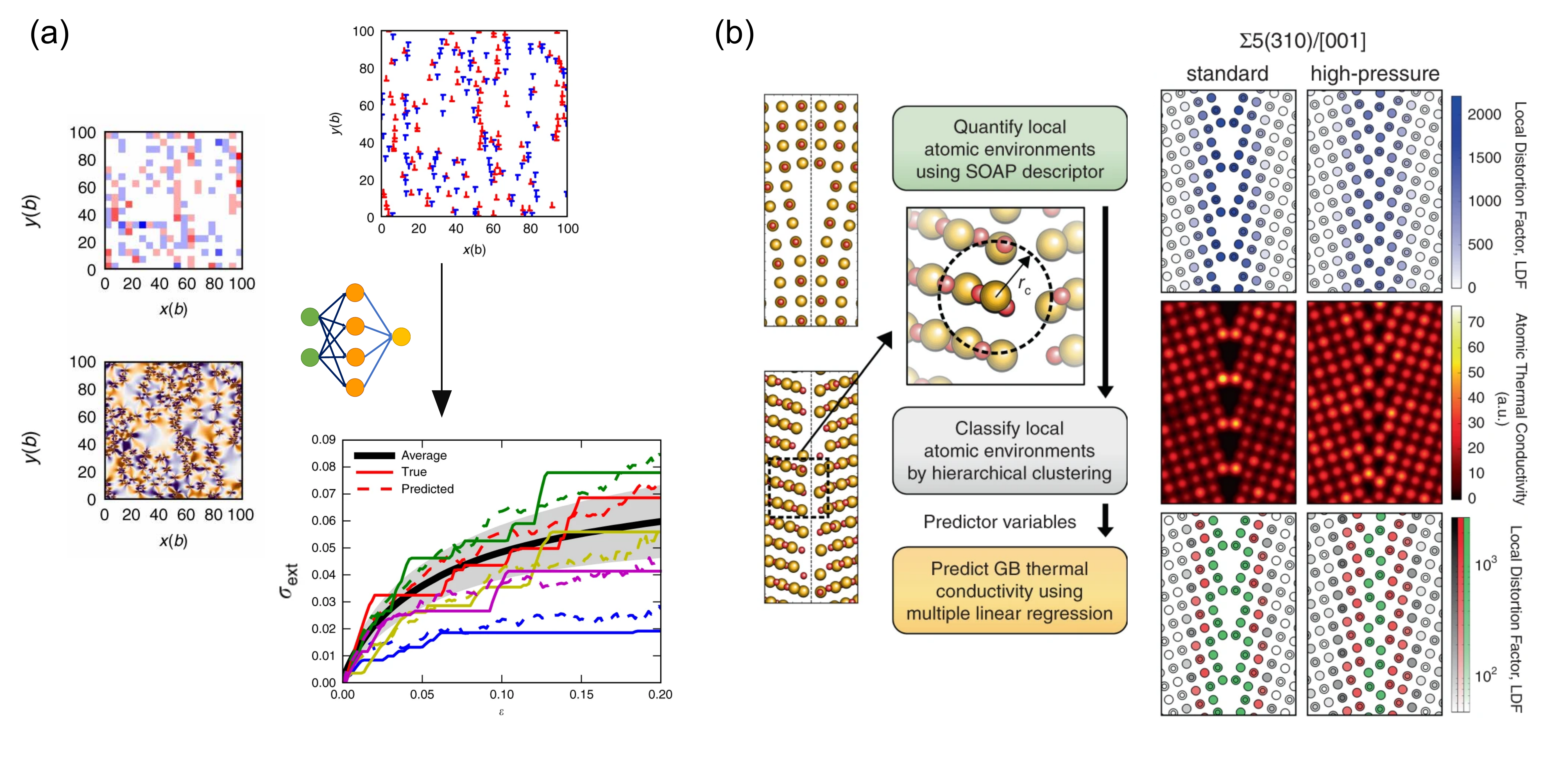}
    \caption{\textbf{ML for the dislocation and the grain boundary}. \textbf{(a)}  The density of geometrically necessary dislocations (left top) and the internal stress field(left down) are taken as the featured descriptor to describe the initial dislocation configuration (right top) and predict the stress-strain curves for plastic deformation. The neural network is trained to infer the relation between features of the initial dislocation configurations and the ensuing stress-strain curves. Reproduction with the permission from Ref.\cite{salmenjoki2018machine} \textbf{(b)} Method for predicting grain boundary thermal conductivities based on local atomic environments. On the left, the SOAP descriptor is utilized for grain boundary structure while $r_c$ is the cutoff radius of the SOAP descriptor. On the right, local distortion factors, Gaussian-smeared atomic thermal conductivities, and distributions of local atomic environments groups classified from hierarchical clustering (green: highly under-coordinated; red: moderately under-coordinated or strongly strained; grey: moderately strained, weakly strained or bulk-like) are presented from top to down to reveal the relation between the grain boundary structure and the thermal transport. Reproduction with the permission from Ref.\cite{fujii2020quantitative}}
    \label{fig:boundary}
\end{figure}

The work observed that the SOAP-constructed LDF correlates properly with atomically decomposed thermal conductivities perpendicular to the grain boundary extracted from perturbed MD simulations. It has been applied to the study of thermal transport to benefit TE based on MD simulations to correlate how the local atomic environment related to the grain boundary affects thermal transport, as shown in Fig.\,\ref{fig:boundary}(b).
Similarly, different descriptors can also be employed to examine the robustness of the ML predictive model and open the way to decouple thermal and electrical conductance
\cite{lortaraprasert2022robust}. This work makes use of five different descriptors to construct the ML model for thermal boundary conductance and electric boundary conductance, the standard deviation of bond angles and bond lengths, the entropy of dihedral angles, and the radial distribution function-related descriptor. Combined with atomic Green's function to produce the thermal and electrical transport data, this work reveals how the structure of grain boundary contributes to phonon and electron transport. Besides, it also indicates that the variations in interatomic angles and distances at the GB are the most predictive descriptors for thermal boundary conductance and electrical boundary conductance, respectively. There are also several other descriptors developed specifically for grain boundaries, as recently summarized in 
\cite{owens2025feature}. This work compares different combinations between descriptors, transforms, and ML algorithms by conducting feature engineering and examining the accuracy and efficiency of the predictive models. Interestingly, this work implies that the incorporation of additional descriptor information can enhance accuracy, although it also introduces extra complexity. Furthermore, some descriptors that encode too much information cannot be easily inverted to extract physical insights from model predictions. All these attempts inspire the community to continue to search for a proper descriptor for the exploration of how to use ML to achieve better TE performance by interface and grain boundary engineering.

Similar to the case of entropy engineering and dislocation, ML force field and interatomic potential also play a significant role in interface and grain boundary-related TE design to generate accurate predictive data efficiently \cite{hu2021perspective,qin2023machine,fujii2022structure,huang2023grain,huang2024unphysical}. The common workflow of these contributions is to construct the ML-based force field or interatomic potential from the DFT-level calculation and then conduct the MD simulation with the grain boundaries and interface. Finally, conduct mostly the thermal transport calculation within the MD simulation and observe how grain boundaries and interface affect the transport behavior. It is also worth noting that some work has indicated that neural network potential demonstrates a remarkable ability to accurately predict low-energy structures and energetics of grain boundaries, achieving reasonable agreement with DFT results \cite{yokoi2022accurate}. In contrast, this work also reveals that conventional empirical potentials fail to identify low-energy grain boundary structures. This reflects the necessity of integrating ML into large-scale simulations to efficiently generate precise and reliable simulated data.

However, current studies using ML potential to simulate large-scale grain boundaries and interfaces focus primarily on thermal transport around the interfaces or grain boundaries, while electrical transport also plays a crucial role of $zT$ \cite{hu2022carrier}. Therefore, a comprehensive approach that integrates the study of both electrical and thermal transport and decouples their effect using ML will be essential \cite{lortaraprasert2022robust}. In addition, as mentioned before, the universal atomic potential or the foundational ML model covering a wide range of elements will further accelerate the simulation for the screening of the high $zT$ grain boundaries and interface design.

Finally, a few examples of applying ML in grain boundary structure studies are discussed. Wagih \textit{et al.} combines SOAP as the descriptor to study grain boundary segregation energy spectra for more than 250 metal-based binary alloys from MD simulations \cite{wagih2020learning}. Recently, this approach has been extended to directly learn segregation energy spectra from first-principles calculations \cite{wagih2022learning}. Similarly, SOAP is also applied to study the atomic energy in grain boundaries \cite{song2022atomic}. The Scattering Transform (ST) is another descriptor also employed to represent grain boundary structures \cite{homer2019machine}. Additionally, it is applied to explore the prediction of grain boundary energy, temperature-dependent mobility, and shear coupling benchmarked with SOAP, demonstrating the effectiveness of these descriptors in capturing the key features of grain boundary behavior. These results will motivate future TE studies on optimizing grain boundaries with ML.

\subsection{Other structural defects}

 The current progress in applying ML to study how key material defects in different dimensions participate in the TE design has been reviewed. Many other forms of structural defects and heterogeneities, such as superlattices, porous structures, precipitates, and embedded quantum dots, have yet to be discussed. In the following discussion, some of these structures are briefly explored to highlight the broader applicability of ML in managing and optimizing various complexities for TE performance.

A superlattice is a periodic structure formed by alternating layers of two or more different materials at the nanoscale. It is engineered to exhibit unique electronic, optical, or thermal properties. The alternating layers introduce extra complexity while providing the tunability of the materials' properties, and the design of these alternating layers can benefit from data-driven methods. Recently, ML has been combined with MD simulations to maximize Anderson localization of phonons within the superlattice system \cite{chowdhury2020machine}. Similarly, by combining with CNN-based ML, people also achieve an active learning framework for aperiodic superlattice design of unexpected thermal transport enhancement with the aid of non-equilibrium MD \cite{roy2022unexpected}. This may come from certain forms of aperiodicity in large-period superlattices that can enable stronger coherent phonon transport that is not localized. The disorder of layer thicknesses of these random multilayer (RML) structures, or aperiodic superlattices, can also be quantified and these disorder parameters can be correlated with thermal conductivity \cite{roy2022unexpected}. The interfacial thermal resistance of the superlattice can also be predicted with ML based on data from the literature \cite{wu2018electrically}. These design strategies should also serve as an inspiration for TE design.

Porous structures can be considered as 3D volume defects in bulk materials. There have been many studies on the enhancement of $zT$ in porous structures \cite{xu2017highly,giulia2017thermoelectric}. ML has also been employed in some TE scenarios involving porous structures \cite{yang2023machine,10.1063/5.0206287}. Due to its large-scale structure, this strategy to manage disorder is also suitable for the optimization of the TE performance at the device level \cite{10.1063/5.0206287}. 

Another important direction is ML-based design for multiple quantum dots systems \cite{zhou2020neural}. Quantum-dot-based hybrids \cite{urban2015prospects} provide a solution to achieve high $zT$ with many specialized features such as their discrete energy levels \cite{amlani1999digital}. This study by Zhou \textit{et al.} explores the use of neural networks to model and optimize the TE properties of multiple interacting quantum dots (MQD). By integrating ML with quantum master equation (QME) simulations, the authors demonstrate that a two-layer NN can effectively predict key transport properties, such as electrical conductance, Seebeck coefficient, and thermal conductance, with significantly reduced computational cost. Similar ML-based approaches can be applied to MQD or precipitation or some other nanostructure-embedded systems designed for TE performance \cite{vineis2010nanostructured,zhao2017superparamagnetic}, where ML can be used to model the interactions between embedded MQD and their surrounding matrix, optimize TE properties, and reduce computational cost in large-scale simulations.

Structural complexity is certainly not limited to what has been discussed above. Other scenarios may include amorphous structures \cite{nolas2002figure}, which share similarities with high-entropy materials. Notably, structural disorder poses an even greater challenge than configurational disorders. However, recent advances have started to explore the application of ML in designing materials with amorphous structures  \cite{liu2023unraveling}. Overall, it is clear that integrating ML into the design process effectively aids in leveraging structural complexity to optimize TE performance.

\section{Outlook}
Despite the extensive efforts in TE engineering, the search for materials with high $zT$ values is just beginning.
There is no universal upper limit to $zT$ in theory, except in specific theoretical models such as the Mahan-Sofo limit \cite{mahan1996best} and its generalization \cite{ding2023best}, and TE materials reach the Carnot limit for energy conversion efficiency at $zT \to \infty$. However, it has been a long-standing challenge to find candidates with even $zT>2$, let alone $zT \gg 1$.
Furthermore, the current progress indicates that the power of AI for TE design is still in its infancy with a large room for improvement, for reasons described below.

\subsection{Curation of extensive database}
There is a shortage of databases for TE materials.
Compared to regular databases for crystal properties, the limited amount of TE datasets arises from two aspects. On the one hand, the current efforts remain insufficient for building a general database of defect configurations. Current defect databases are still limited to $\sim 10^3$ defect configurations \cite{thomas2024substitutional,bertoldo2022quantum,davidsson2021adaq} for specific host materials, and this is far from enough to build a complete understanding of defects in the vast inorganic chemical space.
On the other hand, it is also computationally heavy to calculate \textit{ab initio} transport properties for \textit{crystals}\textemdash let alone materials with defects. 
Boltzmann transport equation (BTE) works well within the diffusive transport regime\cite{madsen2006boltztrap,pizzi2014boltzwann}, but the relaxation time $\tau$ requires phonon calculation of the third-order force constants, which has high computational cost \cite{li2014shengbte,togo2023first,togo2023implementation}. In addition, the non-equilibrium Green's function (NEGF) approach can also be employed to optimize TE at the mesoscopic device level \cite{papior2017improvements,brandbyge2002density}. 

Given these two challenges, there is room for significant improvement in the creation of a more extensive TE database. Future works should focus on systematically incorporating a wider range of defect configurations and host materials. Additionally, more efficient high-throughput DFT workflows are necessary to build the database by performing complicated simulations of transport properties. Integrating these datasets with experimental data will also help overcome current limitations and enable a more comprehensive understanding of TE materials.

\subsection{Search for intrinsic low thermal conductivity}

In the Phonon-Glass Electron-Crystal strategy, while doping and structure defects can be used to tailor electron and phonon transport, the intrinsic low thermal conductivity will be a good starting point for optimizing the TE materials. The intrinsic thermal conductivity of a crystalline material can be reduced by enhanced phonon-phonon scattering from anharmonicity \cite{zhu2017compromise,chang2018anharmoncity,eivari2021low}. This requires the atomic configuration to survey a non-parabolic potential energy profile at the temperature of interest. Enhanced anharmonicity can be achieved through several mechanisms, often by introducing frustration or ambiguity on where an atom prefers to reside or on competing bonding scenarios between neighboring atoms. For example, rattling atoms in cages are usually associated with significant anharmonicity and, therefore, intrinsic low thermal conductivity \cite{tadano2015impact,dutta2024role}. The potential energy profile for the rattling atom is determined by the structure of the cage and the type of the central atom, and the right combination can lead to a highly non-parabolic potential energy well (e.g., flat at the bottom with sharp-rising walls near the boundary). The central atom ``rattles" within such a cage, resulting in significant phonon-phonon scattering. Another example is resonant bonding, where a central atom has an equal opportunity to form covalent bonds with multiple equivalent neighboring atoms. This results in a ``resonance" of bonds between the central atom and all the equivalent neighbors \cite{lee2014resonant,qin2016resonant,zhang2024low}. This complex and non-binary bonding scenario often leads to a highly non-parabolic potential energy profile and thus anharmonicity. Lone pair electrons provide yet another mechanism to reduce intrinsic thermal conductivity \cite{nielsen2013lone}. Because of their deformability, the localized non-bonding lone pair electrons can respond to atomic displacement in a nonlinear way, causing phonon instability and thus anharmonicity. It has been shown that certain crystal structures and element groups are particularly promising in hosting such phenomena \cite{nielsen2013lone}. Apart from the three specific mechanisms mentioned above, there are other more generic strategies to lower intrinsic thermal conductivity, such as introducing heavier elements and weaker bonds and looking for the desired crystal structure with balanced complexity \cite{haque2021first,guillemot2024impact,zeng2024pushing}.

How can ML help to achieve low intrinsic thermal conductivity? First, with an available thermal conductivity database and proper crystalline structure representations, ML models can be trained to establish structure-property relationships. Direct prediction of thermal conductivity from the crystal structure is thus possible \cite{zhu2021charting}, and statistical analysis can be performed to reveal the underlying trend and search for new materials with low thermal conductivity. Second, recent advances in the development of universal machine learning interatomic potentials (MLIPs) can greatly reduce the cost of calculating intrinsic phonon properties, such as anharmonicity, allowing large-scale search and screening of anharmonic crystals. This, however, requires the MLIPs to be sufficiently accurate for calculating higher-order force constants \cite{bandi2024benchmarking}. Most available models so far have yet to be fully benchmarked for such purposes, although tests on harmonic phonons have been reported \cite{deng2025systematic}, with only a few most recent MLIPs showing satisfactory performance \cite{loew2024universal}. Developing or fine-tuning more accurate universal MLIPs \cite{pota2024thermal} for higher-order force constants and anharmonicity is thus a prerequisite for this approach to be successful. Third, one can target a specific mechanism, such as orbit-lattice coupling \cite{wang2024orbit}, and design generative models to extrapolate from known materials to new compositions. 
For instance, certain crystal structures, such as rock-salt, skutterudite, Zintl, and clathrates, have been associated with low intrinsic thermal conductivity. Exploring new recipes (compositions) and synthesizing more compounds of such crystal structures will lead to new TE materials. While generating candidate crystals can be readily achieved with various methods, such as diffusion models, the key challenge here is to down-select the candidates, model/predict their thermal conductivity, and eventually synthesize the material in a laboratory. Multilevel down-selection involving data-driven predictions, universal MLIPs, and conventional DFT screening may be needed to achieve balanced efficiency and accuracy \cite{okabe2024structural}.

\subsection{Inverse design for thermoelectrics}
While ML has demonstrated remarkable success in the forward prediction of TE properties, the inverse design remains in its infancy compared to other areas of materials design \cite{han2024ai,zunger2018inverse}. 
Current workflows primarily rely on forward screening methods, where properties are predicted for pre-defined compositions or structures, followed by labor-intensive optimization. However, this approach is time-consuming and may overlook potentially high-performing candidates.

Inverse design techniques, especially powered by the most recent generative models, offer a promising alternative to accelerate materials discovery. 
For instance, models like MatterGen \cite{zeni2025generative} and CDVAE \cite{xie2021crystal} can perform conditional optimization and sampling, generating candidate materials that meet the desired property criteria. Such models allow researchers to move beyond random or heuristic-based exploration, instead directly targeting regions of the chemical and structural space that are likely to yield optimal performance.
Future advancements in inverse design could incorporate defect-specific conditions to refine TE materials further. By developing such frameworks, the inverse design could revolutionize the discovery process, enabling efficient exploration of vast chemical spaces and more effectively guiding experimental efforts.

\begin{figure}[t]
    \centering
\includegraphics[width=0.7\linewidth]{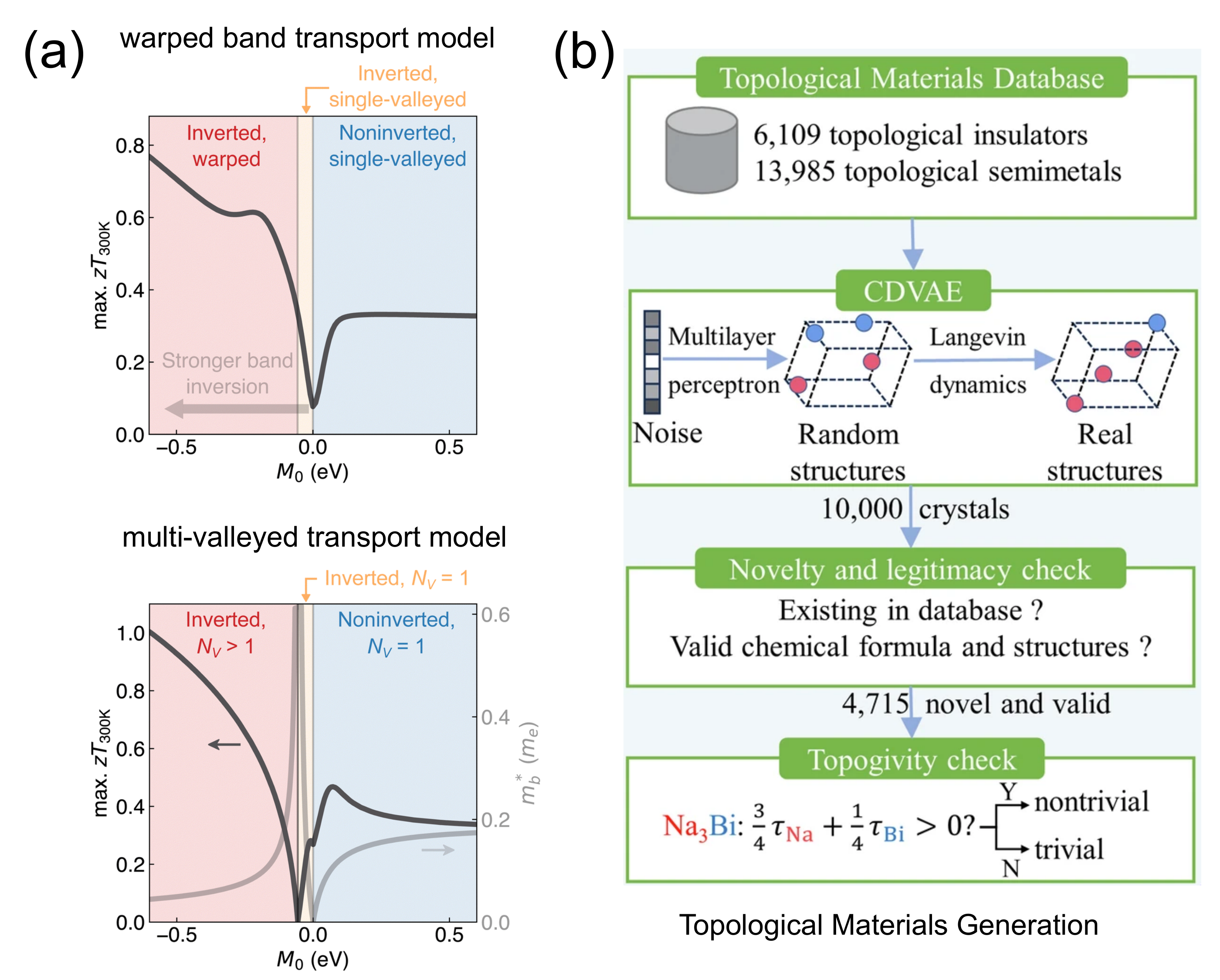}
    \caption{\textbf{Search high $zT$ materials from the topological materials which can be inversely generated with the generative machine learning (ML).} \textbf{(a)} two transport models indicate that strong band inversion will lead to higher $zT$ performance. $M_0$ in the x-axis is a measure of the energy separation of the band edges at the $\Gamma$-point for the present set of materials, where $M_0> 0$ represents noninverted bands (i.e., normal insulators) and $M_0 < 0 $represents inverted bands (i.e., topological insulators) \cite{toriyama2024topological}. \textbf{(b)} The workflow of the generative model for topological materials \cite{hong2025discovery}. The generative model, CDVAE \cite{xie2021crystal}, is trained from the topological materials database, while further filter checks the novelty, legitimacy, and topogivity to screen the targeted topological materials. Reproduced with permission of Ref.\cite{toriyama2024topological,hong2025discovery}.}
    \label{fig:topo}
\end{figure}

\subsection{Machine learning opportunity with novel physical mechanisms}

This review mainly focuses on how to use ML to optimize $zT$ by exploring the structural complexity introduced by defects. This approach aims to integrate defect engineering with ML to enhance TE performance. However, it is important to note that material complexity is not solely limited to defects. Other underlying physical phenomena, such as intricate electronic band structures (e.g., in topological insulators \cite{toriyama2025topological}), hybrid organic-inorganic systems and correlated electron systems that go beyond the traditional Fermi gas model also hold significant potential for improving $zT$ \cite{urban2019new}. ML as a general tool has the potential to address the materials searching or optimization in these cases as well.

One recent promising direction comes from the special electron band, such as the topological materials \cite{han2020quantized}, which demonstrates a quantized TE Hall effect in the Weyl semimetal TaP, enabling high electronic entropy, non-saturating thermopower, and a giant power factor. The gapless $n=0$ Landau level and topological protection enhance TE performance, offering a new pathway for efficient low-temperature energy harvesting. Recently, high-throughput DFT calculations combined with transport modeling also illustrate that some warping driven by band inversion is the key feature for high TE performance in topological insulators \cite{toriyama2024topological}. In this work, different transport models are studied to reveal that stronger band inversion may lead to $zT$ higher than normal insulators with noninverted bands, as shown in Fig.\,\ref{fig:topo}(a). ML may play a role in the search for topological materials with specialized band structures to benefit the possible enhancement of TE performance. For example, some generative models \cite{hong2025discovery} have been used to discover new topological materials with further consideration of TE performance. In this generative model, the topogivity is taken to determine whether a material is topological or not \cite{ma2023topogivity}. 

Similarly, hybrid organic-inorganic materials have recently attracted more attention for potential TE applications. A distinct feature of hybrid materials, compared to composites or material mixtures, is that the overall $zT$ of the hybrid organic/inorganic system can be higher than either individual component \cite{urban2019new} due to the physical interactions between components. The new phases of materials that conjoin inorganic and organic materials have shown simultaneous improvement in $S$ and the $\sigma$ \cite{cho2015engineering}. When combined with ML, this kind of hybridization opens up a large exploration space to help the future design of TE materials.

In all these cases, research is still in its early stages. However, the integration of novel mechanisms for enhancing TE performance with ML will unleash its full potential, driving the accelerated discovery and development of advanced TE materials.

\subsection{Thermoelectric materials with sustainability}

Opportunities with ML are not limited to the areas mentioned above. On top of tackling technical challenges originating from the complexity of TE materials, the potential of ML can also be expanded to develop efficient and environmentally sustainable energy systems \cite{yao2023machine}. In the past, enhancing material performance was the primary focus of material design and engineering research, where sustainability had not been considered.  Nevertheless, sustainability is an essential criterion for assessing material’s feasibility and reliability for applications outside the laboratory environment.

\begin{figure}[t]
    \centering
\includegraphics[width=0.9\linewidth]{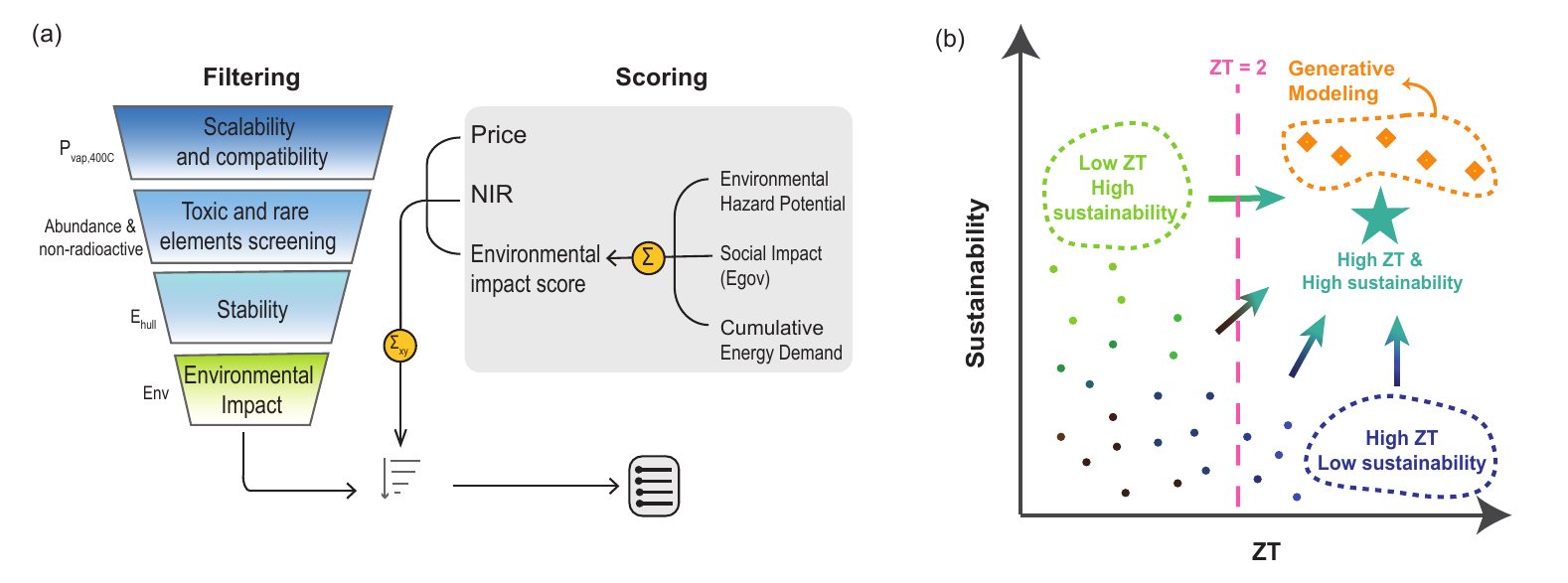}
    \caption{\textbf{Consider sustainability in future TE design.} \textbf{(a)} Down-selection and scoring of topological materials. The selection is limited to stable materials
 (E$_{hull}<$ 0) characterized by non-toxic elements, a modest melting temperature T$_{melt}<$ 2000 K, and low vapor pressure. These materials can then be sorted based on price, net import resilience (NIR), and environmental impact score (Env). When applied to the topological materials database, this process yields roughly 200 materials candidates from a pool of over 16,000. Reproduced with permission from Ref.\cite{boonkird2023sustainability}. \textbf{(b)} The paradigm of
 inverse design with generative modeling pushes the material space to overcome the performance and sustainability dilemma, figure adapted and get permission to reproduce from the \cite{cheng2025ai}.}
    \label{fig:sustainable}
\end{figure}

However, evaluating sustainability is an inherently complex challenge due to the lack of standardized metrics. The definition of sustainability can vary depending on the context. For example, in the context of sustainable energy, sustainability implies low or zero carbon emission, stable source, or cost of electricity generation. From the perspective of the materials, key considerations include abundance, stability, durability, and environmental impact, which encompasses factors such as toxicity, resource extraction, fabrication, disposal costs, and recyclability. To assess the sustainability of each material, factors influenced by social variables, such as market price, infrastructure, and policy, must also be considered, making it challenging to translate the costs of these elements during the material development stage. 

As such, due to the various meanings implied by sustainability, there is a need to quantify it into indicators that can be analyzed using objective criteria. Recently, several initial data-driven attempts have been made to bring together all sustainability-related factors and perform filtering and scoring to recommend topological materials \cite{boonkird2023sustainability}. The selection is limited to stable materials characterized by non-toxic elements, a modest melting temperature, and a low vapor pressure, and then the selection is sorted based on price, net import resilience (NIR), and environmental impact score (Env) as shown in Fig.\,\ref{fig:sustainable}(a). This down-selection and scoring search can be conceptualized as the recommendation system and further developed through modern ML algorithms \cite{zhang2019deep}. Another way to bridge the gap between materials sustainability and performance with the ML method is to utilize the conditional generative model \cite{cheng2025ai}. With sustainability as a guiding constraint, the model can generate a large number of candidate materials, enabling the search for optimal materials that balance the competing factors as shown in Fig.\,\ref{fig:sustainable}(b). This approach has the potential to lead to breakthrough innovations in the energy materials ecosystem.

\section*{Acknowledgement}
CF and MC acknowledge support from the U.S. Department of Energy (DOE), Office of Science (SC), Basic Energy Sciences (BES), Award No. DE-SC0021940. RO acknowledges support from the National Science Foundation (NSF) Designing Materials to Revolutionize and Engineer our Future (DMREF) Program with Award No. DMR-2118448. ML acknowledges the Class of 1947 Career Development Chair and the support from R. Wachnik. YC was supported by the Scientific User Facilities Division, BES, DOE, under Contract No. DE-AC0500OR22725 with UT Battelle, LLC.

\section*{Data availability}
This review does not involve the generation of new data.

\section*{Conflict of interest}
The authors declare no conflict of interest.

 \bibliographystyle{elsarticle-num} 
 \bibliography{cas-refs}





\end{document}